\documentclass[12pt]{article}
\usepackage{graphicx,amsmath,amssymb}
\usepackage[sort]{cite}
\usepackage{graphicx}
\usepackage{multicol}
\usepackage{xcolor}
\usepackage{undertilde}

\def\lsim{\mbox{\raisebox{-.5mm}{$\,\stackrel{<}{\scriptstyle\sim}\,$}}}

\newcommand{\be}{\begin{equation}}
\newcommand{\ee}{\end{equation}}
\newcommand{\bea}{\begin{eqnarray}}
\newcommand{\eea}{\end{eqnarray}}
\begin{document}

\thispagestyle{empty}
\vspace*{.2cm}
\noindent
 \hfill 9 April 2013

\vspace*{2.0cm}

\begin{center}
{\Large\bf The Higgs mass from a String-Theoretic Perspective}
\\[2.5cm]

{\large Arthur Hebecker$^1$, Alexander K.~Knochel${}^{1,2}$ and Timo Weigand$^1$\\[6mm]}

{\it
$^1$Institut f\"ur Theoretische Physik, Universit\"at Heidelberg, 
Philosophenweg 19,\\ D-69120 Heidelberg, Germany\\[1mm]
$^2$Institut f\"ur Theoretische Teilchenphysik und Kosmologie, RWTH Aachen, \\
D-52056 Aachen, Germany\\[3mm]

{\small\tt (\,a.hebecker} {\small,}
{\small\tt \,t.weigand@thphys.uni-heidelberg.de}\small\tt \,{\small;} knochel@physik.rwth-aachen.de) }
\\[2.0cm]

{\bf Abstract}
\end{center} 

\noindent
The Higgs quartic coupling $\lambda$ has now been indirectly measured at the 
electroweak scale. Assuming no new low-scale physics, its running 
is known and, together with gauge and Yukawa couplings, it is a crucial
new piece of information constraining UV completions of the Standard Model. 
In particular, supersymmetry broken at an intermediate or high energy scale with 
$\tan\beta=1$ (i.e. $\lambda=0$) is consistent with present data and has an 
independent theoretical appeal. We analyze the possible string-theoretic 
motivations for $\tan\beta=1$ (including both the shift-symmetry and the 
more economical variant of a $Z_2$ symmetry) in a Higgs sector realized on 
either 6- or 7-branes. We identify specific geometries where $\lambda 
\simeq 0$ may arise naturally and specify the geometrical problems which need
to be solved to determine its precise value in the generic case. We then analyze 
the radiative corrections to $\lambda$.  
Finally we show that, in contrast to naive expectations,
$\lambda<0$ at the SUSY breaking scale is also possible. Specifically, string
theory may produce an MSSM plus chiral singlet at a very high scale, which
immediately breaks to a non-SUSY Standard Model with $\lambda<0$. This
classically unstable theory then becomes metastable through running towards the
IR.

\newpage
\section{Introduction} It is now established that the recently
discovered resonance described  in \cite{expatlas,expcms} is at least very 
similar\cite{Plehn:2012iz} to the
Standard Model Higgs boson. Together with constraints from electroweak
precision data, flavor physics (see e.g. \cite{Baak:2011ze,Isidori:2010kg} and
refs. therein) and quickly evolving direct-search limits from LHC, this
supports the conservative hypothesis that the Standard Model remains the
correct effective field theory above the TeV scale. To the best of our
knowledge, this implies that the electroweak scale is fine-tuned, which may
indeed be acceptable in view of the huge landscape of string vacua (see e.g.
\cite{Denef:2004ze}). 

Thus, we focus on string theory as {\it the} high-scale theory and, more
specifically, on those solutions of the string equations of motion which can be understood as compactifications of 10d
supergravity (F-theory models, heterotic Calabi-Yaus, or type II Calabi-Yau
orientifolds with branes).\footnote{ Of course, entirely stringy constructions
without any higher-dimensional interpretation and with supersymmetry broken at
the string scale are also conceivable. The pragmatic reason for not considering
them here is our lack of quantitative understanding. A possible physics reason
is their short expected life-time: More or less by definition, there is no
small parameter controlling tunneling rates in the corresponding part of the
landscape.  } For reasons of stability, we expect supersymmetry at the
compactification scale $m_C$. If we accept fine tuning in the Higgs sector,
SUSY can then be broken anywhere below that scale, $m_Z\ll m_S\leq m_C$. For
the purpose of this paper, we mostly focus on the `canonical' minimal particle content
at $m_S$: a high-scale MSSM. We furthermore assume that all superpartner masses
are of the same order of magnitude (see however \cite{ArkaniHamed:2004fb}),
such that $m_S$ is reasonably well-defined. 

Now, given the measured Higgs mass, the electroweak-scale quartic coupling
$\lambda$ of the Higgs field and its running at higher energies are in principle known. In
particular, $\lambda$ runs to zero at some high scale
$\mu_\lambda$\cite{Lindner:1988ww,Shaposhnikov:2009pv,Cabrera:2011bi,Degrassi:2012ry,Holthausen:2011aa,Giudice:2011cg,EliasMiro:2011aa,Masina:2012tz} (see \cite{threeloop}
for state-of-the-art SM beta functions).
The standard MSSM formula 
\be
\lambda(m_S)=\frac{g_2^2(m_S)+g_1^2(m_S)}{8}\cos^2 2\beta\,
\label{lambda}
\ee
expresses the quartic coupling in terms of the MSSM $\beta$-angle and the electroweak gauge couplings $g_1$ and $g_2$ \cite{Martin:1997ns}.
One sees that $\lambda(m_S)\geq 0$, implying $m_S \leq \mu_\lambda$.
Furthermore, if an argument for a particular value of 
$\tan\beta$ can be made, Eq. (\ref{lambda}) together with the running of $\lambda$ can be used to predict $m_S$.\footnote{ For a detailed numerical study and plots
illustrating this relation see e.g.  \cite{Degrassi:2012ry}.  } This was
exploited in \cite{Hall:2009nd}, with focus on $\tan\beta\gg 1$ at the GUT
scale (now ruled out by data). If, by contrast, a convincing theory reason for
the opposite limit, $\tan\beta=1$ and thus $\cos^2 2\beta=0$, can be provided, high-scale SUSY is
consistent with a 126 GeV Higgs and the SUSY breaking scale $m_S$ coincides with the scale $\mu_\lambda$ at which the quartic coupling $\lambda$ vanishes.
In this scenario, introduced in  \cite{Hebecker:2012qp}, the prediction 
$m_S\sim \mu_\lambda \sim10^9$ GeV (at present still with very large errors)
can be made. More specifically, in
\cite{Hebecker:2012qp} a string-motivated
\cite{LopesCardoso:1994is,Brignole:1997dp} shift symmetry was suggested as the
origin of $\tan\beta=1$.\footnote{
An 
earlier, closely related idea is to realize the Higgs as a pseudo-Goldstone boson \cite{Inoue:1985cw}. Furthermore, in the context of predicting (or explaining) the Higgs mass, non SUSY-variants of a shift symmetry in the Higgs sector appeared several years ago in \cite{Gogoladze:2007qm} as well as very recently in \cite{Redi:2012ad}. A shift-symmetric Higgs has also been discussed in the context of inflation in \cite{BenDayan:2010yz}. An alternative origin of $\lambda(m_S)=0$ was suggested in \cite{Unwin:2012fj}.
}
As observed in
\cite{Ibanez:2012zg} (and further analyzed phenomenologically in
\cite{Ibanez:2013gf}), the weaker assumption of a $Z_2$ symmetry is in fact
sufficient to ensure $\tan\beta=1$. 

There is, however, an interesting alternative
to Eq. (\ref{lambda}) in non-minimal SUSY models which allows $\lambda<0$ at
the soft scale. 
Note that this is independent of the details of the stringy
or supersymmetric UV completion. We will introduce this scenario in section \ref{sec:negativelambda} and argue that an energy window {\it without}
even a metastable vacuum at small or zero Higgs vev might exist. Rather than
extending the SM by a stabilizing sector below $\mu_\lambda$ \cite{EliasMiro:2012ay,Giudice:2011cg}, 
we simply accept that $\lambda$ runs negative and only ensure the existence of
some UV completion at much higher energies.

In the present paper, we attempt to flesh out the possible stringy 
symmetries in the Higgs sector (either shift or $Z_2$) which were
suggested in \cite{Hebecker:2012qp,Ibanez:2012zg} as the origin of 
$\tan\beta=1$. While the shift symmetry is well-understood in heterotic
orbifold models \cite{LopesCardoso:1994is,Brignole:1997dp}, its possible 
generalization to heterotic Calabi-Yaus appears to be complicated 
and we postpone this issue. Instead, we focus on 6- and 7-brane models 
in type IIA and IIB Calabi-Yau orientifolds and in F-theory. 
We also compare shift symmetry violating effects arising from the inequality
$m_C>m_S$ and from loop corrections  at $m_S$.

Before giving an outline of the paper, we now pause to discuss critically 
possible motivations for high-scale SUSY with $\tan\beta=1$. The simplest
motivation, already emphasized above, is that it may relate a symmetry
feature of the high-scale theory ($Z_2$ or shift symmetry) to the last 
observable of the Standard Model (Higgs mass or equivalently quartic coupling
$\lambda$). As such, scenarios with $\tan\beta=1$ are certainly worthy of 
investigation. 

Furthermore, it is conceivable that a better understanding of `how to find 
a Standard Model in the landscape' will provide us with a first-principle 
reason for a symmetry enforcing $\tan\beta=1$. As will become more apparent
below, this is far from obvious. In fact, at the moment it appears that the 
required symmetries are only available in a certain class of models and that
(as usual) many models {\it without} this feature exist. 

Finally, there might be a `landscape bias' towards a high SUSY breaking 
scale. Given a certain low-scale value of $\lambda$ with a corresponding 
`vacuum stability scale' $\mu_\lambda<M_{\rm P}$, this implies a bias for 
$m_S$ to be as close to $\mu_\lambda$ as possible\footnote{Note that there
is an unconventional alternative if one is willing to accept $\lambda<0$ at the
matching scale and possible metastability of the SM vacuum. This possibility is
discussed in section
\ref{sec:negativelambda}. }. In such a situation, models ensuring $\tan\beta=1$ by
a symmetry are very strongly 
favored. While there are obviously many uncertainties in this argument
(e.g. the preferred SUSY breaking scale as well as the whole concept of a 
measure in the landscape), we still believe that it is an interesting extra 
reason to consider the present class of models.

The paper is organized as follows: In the remainder of the introduction, we
briefly recall the role of shift and exchange symmetry in determining the UV
boundary conditions of the Higgs sector masses and couplings.

\noindent
In Section \ref{sec:typeIImodels}, we discuss possible stringy realizations of
shift/exchange symmetric Higgs sectors in smooth Calabi-Yau compactifications.
In compactifications with D-branes, the Higgs field can either propagate along the entire brane as a so-called 
bulk Higgs, or localize at the the intersection of two D-branes.
We argue that a bulk Higgs on D6 branes or its type IIB/F-Theory dual might be
the most natural way to obtain shift symmetric scenarios in type II models.  
Upon deforming parallel branes relative to each other, bulk degrees of freedom become localized matter.
We take a closer look at the fate of shift/exchange symmetry during the
transition from D7 bulk fields to matter localized on the intersection curve of two D7-branes, and identify the point in moduli
space where the transition between the two regimes occurs. We consider simplified
settings in $D=5,6$ with modulus dominated or Scherck-Schwarz supersymmetry
breaking and then draw some conclusions for Calabi-Yau compactifications with
fluxes.

\noindent
In Section \ref{sec:pterm},  we discuss the Higgs quartic coupling in models
where extended SUSY effects can play a significant role. In particular, we are
interested in the recently proposed exchange symmetric models \cite{Ibanez:2012zg,Ibanez:2013gf} on D6 brane
networks exploiting the possibility of relative extended ${\cal N}=2$ supersymmetry preserved by pairs of D6 brane stacks. 
We give field-theoretic arguments that there is some tension between
the need to generate a sizable $B\mu$ term via branes at angles or fluxes, and
the desire to have an MSSM-like $D$-term potential. We discuss variations of the
branes-at-angles scenario in the flat space limit and propose a possible
solution.

\noindent
Section \ref{sec:corrections} addresses both tree level and loop corrections to
the quartic Higgs coupling and the intriguing possibility of negative $\lambda$
at the soft scale. In Section \ref{sec:radcor}, our previous analysis of
radiative shift symmetry violation is updated and augmented by a detailed
treatment of 1-loop threshold effects. We discuss the relative importance of
the various contributions and their impact on the predicted SUSY breaking
scale.  In Section \ref{sec:negativelambda}, we argue that some high-scale UV
completions with approximately flat directions (e.g. approximately shift- or
exchange symmetric models) can yield negative effective quartic couplings and
nevertheless remain calculable perturbatively due to the local flatness of the
potential at small field values. We show how this relaxes the upper bounds on
the SUSY breaking scale, which might be useful for certain classes of string
models (see e.g. \cite{Cicoli:2013rwa}).

\subsection{The role of shift symmetry and exchange symmetry}
\label{sec:tanbeta}
\label{sec:shiftexchangesymm}

Let us review the role of the shift and the exchange symmetries in the UV boundary conditions of the
Higgs quartic couplings before we discuss superstring/F-Theory realizations.

\paragraph{Shift symmetry:}
 We consider an MSSM-like Higgs sector with two chiral Higgs doublets $H_u$ and
$H_d$ carrying opposite hypercharge. We demand that the K\"ahler potential
exhibit a shift symmetry 
\begin{equation}
H_u \longrightarrow H_u + c,\,\quad H_d \longrightarrow H_d - c^\dagger\,.
\end{equation}
The consequence of this symmetry is that the K\"ahler potential can only depend
on the linear combination, $K = K(H_u + H_d^\dagger)$. The
lowest-dimension allowed operator in the Higgs fields is
\begin{equation}K=f(X_i,\overline X_i) |H_u + H_d^\dagger|^2\,,
\label{eq:genericshiftkaehler}
\end{equation} 
where we have included a generic dependence on some moduli $X_i$ which will
later acquire an $F$-term vev.  Furthermore, the dominant source of shift
symmetry violation should come from the couplings, i.e. the quadratic part of
the Higgs lagrangian will be affected by this violation at the 1-loop level
(see section \ref{sec:radcor} for a discussion of these effects).  What is
remarkable about this K\"ahler potential is that it contains the usual kinetic
terms $|H_u|^2$, $|H_d|^2$ as well as a ``holomorphic'' part $H_u H_d +
 H_d^\dagger H_u^\dagger$. The latter by itself would only contribute a
total derivative to the action, but this changes in the supergravity setting,
and specifically due to the nontrivial dependence of $f$ on the moduli.  

In modulus dominated SUSY breaking, some of the moduli $X_i$ will acquire
$F$-term vevs $F^{Xi}$, which in turn can generate soft terms as well as
supersymmetric masses wherever they occur in the K\"ahler potential and
superpotential. In absence of a $\mu H_u H_d$ term in the superpotential, the
Higgs mass matrix can be read off the K\"ahler potential in Eq.
(\ref{eq:genericshiftkaehler}).  Without loss of generality, we take $f=1$ in the
vacuum.  Upon supersymmetry breaking, soft masses $m_{H_u}^2, m_{H_d}^2$, the
$B \mu$ term and the effective $\mu$ term are generated.  For simplicity we now
assume one modulus $X$ to dominate. The resulting Higgs mass matrix (with
$B\mu\equiv m_3^2$) at the scale $m_S$, defined through
\be
{\cal L}\supset -m_1^2 |H_u|^2-m_2^2|H_d|^2-m_3^2 (H_u {H}_d+
{H}_d^\dagger H_u^\dagger)\,,
\ee
is then given by (cf. e.g.~\cite{LopesCardoso:1994is,Brignole:1997dp})
\be
m_1^2=m_2^2=m_3^2=\left|\mu\right|^2+m_{3/2}^2-F\lefteqn{\phantom{\overline{F}}}^X \overline{F}^{\overline{X}}
(\ln f)_{X\overline{X}}\,,
\label{mi}
\ee
where 
\be
|\mu|^2=\left|m_{3/2}-\overline{F}^{\overline{X}}
f_{\overline{X}}\right|^2\,,\qquad
\overline{F}^{\overline{X}}=e^{K/2}K^{\overline{X}X}D_SW\qquad\mbox{and}\qquad
m_{3/2}=e^{K/2}W\,.
\ee
As was discussed in \cite{Hebecker:2012qp}, this Higgs mass matrix has a
massless eigenstate 
\be
H_0 =\frac{1}{\sqrt{2}}(H_u-H_d^\dagger)\,.
\label{masslesshiggs}
\ee
It provides the SM Higgs doublet, while the orthogonal combination becomes
heavy at the soft scale and does not contribute to EWSB.  This situation
corresponds to the decoupling limit with a mixing angle $\tan\beta=1$. Since we
get a zero eigenvalue, it appears as though we get a naturally low  (compared
to the soft scale) electroweak scale and have avoided the hierarchy problem.
Unfortunately, this changes once one goes beyond tree level. After integrating
out heavy degrees of freedom coupling to the Higgs sector like the top squarks,
the corresponding decoupling contributions, though loop suppressed, give us
large shifts in the Higgs mass parameters which force us to finetune the
electroweak scale. The fact that these contributions are loop suppressed with
respect to the treelevel mass parameters ensures that $\tan\beta \approx 1$
still holds to good accuracy after loop corrections are taken into
account. 

One might worry that the shift-symmetric scenario is too restrictive to tune
the Higgs light after threshold corrections at $m_S$ are included. However,
since the shift-symmetry violating contributions themselves are enhanced
relative to the quadratic threshold corrections by a $\log m_S/m_C$, one can
tune the Higgs light within the leading log approximation by dialing the soft
terms without having to give up the shift-symmetric scenario
\cite{Hebecker:2012qp}.

\paragraph{Exchange symmetry:}
One can ask whether there are weaker conditions than a shift-symmetric Higgs
sector which also yield the desired structures. In \cite{Ibanez:2012zg}, the
authors find that there is a simpler way to enforce $\tan\beta=1$ by symmetry
if one does not insist on a naturally small electroweak scale at tree level.
Since one needs to finetune anyways after taking into account quadratic loop
corrections, we do only a factor of $\approx 6 y_t^2/16 \pi^2$  worse if we
give up this requirement altogether, as was also pointed out in
\cite{Ibanez:2013gf}.  The idea is then as follows: if the Higgs potential
exhibits an exchange symmetry (again, necessarily only at tree level) between
$H_u$ and $H_d^\dagger$, this ensures that the diagonal entries in the Higgs
mass matrix satisfy $m_{H_u}^2=m_{H_d}^2=m^2$, and thus after canonical
normalization, it becomes
\be
M(m_S)=
\left[
\begin{array}{cc}
|\mu|^2+m^2 & B\mu  \\
B\mu  & |\mu|^2+m^2
\end{array}
\right]\,.\ee
The electroweak scale is tuned light when
\begin{equation}B\mu\approx|\mu|^2+m^2\mbox{+quadratic thresholds}\,.\end{equation}
Thus, the requirement of having a light electroweak scale already implies that
$m_3\approx m_1=m_2$, and we recover the same Higgs mass matrix with universal
entries as before and thus $\tan\beta\approx 1$ along with a massless doublet.
If the relation $m_{H_u}^2 = m_{H_d}^2$ receives small corrections, the
resulting deviation from $\tan\beta=1$ and thus $\cos 2\beta=0$ is
\begin{equation}
|\cos2\beta| =\frac12 \frac{|m_{H_u}^2 - m_{H_d}^2|}{|B \mu|}
\end{equation}
at leading order in $\Delta m^2/\mu^2$. In a sense, the shift-symmetric
scenario is a special case of the exchange symmetric one, and both symmetries
are equally broken by loop effects from terms such as the large third
generation Yukawa coupling, which for $\tan\beta=1$ is only present for $H_u$,
but not $H_d$. Our treatment of radiative corrections in section
\ref{sec:radcor}, which is an extension of the analysis given in
\cite{Hebecker:2012qp}, is thus also valid for this case. This is also in
agreement with a recent analysis in \cite{Ibanez:2013gf}. A particularly simple
version of the exchange symmetric scenario is realized if the soft masses
$m_{H_u},m_{H_d}$ in the Higgs sector vanish altogether or are strongly
suppressed relative to $|\mu|^2$. In \cite{Ibanez:2012zg}, such a scenario is proposed
using brane angles or open string fluxes to generate a $B\mu$ term.  Another possible
realization of this idea, which we will not consider further in this work,
assumes that supersymmetry breaking is communicated to the Higgs sector via a
renormalizable interaction $\mathcal W \sim SH_u H_d$ with a  ``PQ-spurion''.
If $\langle S\rangle =s+\theta^2 F^S$, the resulting $\mu/B\mu$ terms, which
yield exchange symmetric normalized masses, might dominate over
exchange-violating soft masses originating from higher-dimensional operators.
The Higgs mass matrix then has the desired approximate exchange-symmetric form,
and the electroweak scale can be tuned light by varying $\mu_{eff}$ versus
$B\mu$.  However, the scalar $F$-term will also contribute to the quartic Higgs
interaction at tree level if $S$ is not decoupled in an approximately
supersymmetric fashion above the scale $m_S$, and this can spoil the Higgs mass
predictions from $\tan\beta=1$ (see section \ref{sec:pterm}). Also, if this
mechanism yields large hierarchies between the heavy Higgs doublet and the
sfermions or gauginos, the threshold corrections to $\lambda$ may be the
dominant effect.

\section{Type IIB/F theory with 7-branes}
\label{sec:typeIImodels}
The earliest and most explicit string-theoretic models with a shift-symmetric Higgs doublet pair
are heterotic orbifolds \cite{LopesCardoso:1994is, Buchmuller:2005jr}.  As
explained in some detail in \cite{Choi:2003kq,Brummer:2010fr, Hebecker:2012qp}, the shift
symmetry emerges since the Higgs is a Wilson line\cite{Ibanez:1987xa} (see \cite{Pena:2012ki} for a recent
discussion of twisted vs. untwisted Higgs sectors in heterotic orbifolds). 
Here, we
are interested in generic, smooth Calabi-Yau compactifications. This
generalization is problematic \cite{Hebecker:2012qp} since, in going from
orbifold to Calabi-Yau, the Higgs becomes one of the bundle moduli and the
corresponding moduli space is less well understood. 

\subsection{Shift symmetry: Wilson-lines vs. brane-deformations}

We therefore turn to brane models, for concreteness type IIB with D7-branes (see e.g. \cite{Aldazabal:1998mr} for constructions on orbifolds and smooth Calabi-Yau manifolds),
where the Higgs once again has the chance of being simply a Wilson line.  A
strongly simplified version of the corresponding K\"ahler potential 
\cite{Jockers:2004yj} is
\be
K=-\ln[-i(S-\overline{S})-L(z,\overline{z})\zeta\overline{\zeta}]
  -3\ln[T+\overline{T}-C(z,\overline{z},\zeta,\overline{\zeta})a\overline{a}]
  -K_{cs}(z,\overline{z})\cdots\,.
\label{kaehler2b}
\ee
Here $S$ is the axio-dilaton and $z,\zeta,T,a$ are the complex structure,
D7-brane, K\"ahler and Wilson line moduli respectively. Merely for notational
simplicity, we suppress indices pretending that there were just one modulus of
each kind. 

Since this K\"ahler potential has only been obtained in a fluctuation-expansion
around a given brane configuration, we a priori do not know how the leading
terms $\zeta\overline{\zeta}$ and $a\overline{a}$ generalize to all-orders
K\"ahler metrics on the brane-deformation moduli space and the Wilson-line
moduli-space respectively. 

Naively, one might hope that $a\overline{a}$ actually arises from a term $\sim
(a+\overline{a})^2$ or, more generally, from a {\it shift-symmetric} Wilson
line K\"ahler potential
\be
k_w(z,\overline{z},\zeta,\overline{\zeta},a+\overline{a})\,\,=\,\,
C(z,\overline{z},\zeta,\overline{\zeta})\,a\overline{a}+\cdots\,.
\ee
However, a closer look reveals that things can not be so simple: Our complex
variable $a$ combines two independent Wilson lines corresponding to its real
and imaginary part. Respecting a shift-symmetry in both of them would imply
that $k_w$ should be independent of both $(a-\overline{a})$ and
$(a+\overline{a})$, which is clearly impossible. Indeed, upon closer inspection
of the dimensional reduction carried out in \cite{Jockers:2004yj} one sees that
the Chern-Simons term contains (very schematically) a piece
\be
\int_{D7}C_4 \wedge dA_1 \wedge dA_1 = \int_{D7}A_1 \wedge dC_4 \wedge dA_1\,,
\ee
which induces kinetic mixing between D7 Wilson lines and $C_4$ scalars, with a
prefactor depending on the Wilson-line. This destroys the shift symmetry (at
least generically). 

Of course, even in the absence of a shift symmetry, an exchange symmetry of the
type suggested in \cite{Ibanez:2012zg,Ibanez:2013gf} might still apply. Thus,
we by no means rule out D7-brane Wilson-line scalars as candidates for a Higgs
sector with $\tan\beta=1$. However, we will not develop this line of thinking
in the present paper. 

Next consider the brane deformation moduli $\zeta, \overline{\zeta}$ where, as
we will argue, a shift symmetry does indeed exist (cf. \cite{
Hebecker:2012qp,Hebecker:2012aw,Arends}). To make our point, we consider the
type IIA mirror version of (\ref{kaehler2b}). The analogue of the first term,
again strongly simplified, is \cite{Kerstan:2011dy}
\be
K=-\ln[-i(S-\overline{S})-Q(t,\overline{t})u\overline{u}]+\cdots\,,
\label{kaehler2a}
\ee
where $t$ and $u$ are the K\"ahler and D6-brane moduli respectively. To avoid
losing focus, we do not discuss the type IIA analogue of $S$ in any detail. We
only note that the type IIA axio-dilaton contains both $g_s$ and the volume and
also has an intimate relation to the complex structure through the $C_3$ scalar
(its imaginary part) associated with the holomorphic 3-form \cite{Bohm:1999uk}.
This will not be important for us and we do not go into the details of how $S$
is expressed in terms of IIA compactification data. The claim that such a
superfield exists and the K\"ahler potential can be written as in
(\ref{kaehler2a}) can be viewed as the assertion that mirror symmetry continues
to hold in the context of ${\cal N}=1$ orientifolds.

The main point for us is that the complex field 
\be
u=\Phi+i(A-M\Phi)
\ee
combines one real brane-deformation $\Phi$ and one Wilson line modulus $A$ ($M$
vanishes for vanishing $B_2$ background and will not be relevant for us). In
particular, Re$(u)$ is independent of $A$. Furthermore, as can be seen from
\cite{Kerstan:2011dy}, the dimensionally reduced 4d action involves $A\sim
(u-\overline{u})$ only through its derivatives. This crucially depends on the
structure of the relevant Chern-Simons term,
\be
\int_{D6}C_3 \wedge dA_1 \wedge dA_1 = \int_{D6} A_1 \wedge dC_3\wedge dA_1 \,.
\ee
In contrast to the D7-brane case discussed earlier, this term does not give
rise to any (potentially $A$-dependent) kinetic mixing between the Wilson line
scalar and a $C_3$ scalar. The point is that, for such a mixing term, both
$A_1s$ should be integrated along 1-cycles of the D6. For a non-zero result,
one would need to dimensionally reduce $dC_3$ to a 4d 3-form using a harmonic
Calabi-Yau 1-form, to be then pulled back to the brane. Since Calabi-Yaus have
no 1-cycles, no such kinetic mixing terms arise. 

Thus, we can generalize (\ref{kaehler2a}) by writing 
\be
K=-\ln[-i(S-\overline{S})-k_{D6}{(t,\overline{t},u+\overline{u})}]+\cdots\,,
\ee
or, returning to the IIB side,
\be
K=-\ln[-i(S-\overline{S})-k_{D7}(z,\overline{z},\zeta+\overline{\zeta})]+
\cdots\,.
\label{kd7}
\ee
 Note that we need to be at large complex
structure on the type IIB side in order to trust the classical K\"ahler potential found on the IIA side at large
volume.
This is the desired shift-symmetric structure we want to start from.

\subsection{Bulk Higgs vs. intersection-curve Higgs}\label{bhvs}

Promoting our D7-brane to an $SU(6)$ stack, $\zeta$ becomes an $SU(6)$ adjoint.
After further symmetry breaking to $SU(5)$ (or directly to the Standard Model),
it can then contain the ${\bf 5}+\overline{\bf 5}$ (or ${\bf 2}+\overline{\bf
2}$) Higgses, with a shift symmetric K\"ahler potential. Models of this type
have recently been considered in F-theory \cite{Donagi:2011dv}. For reviews
of model building in IIB/F-theory see e.g. \cite{Maharana:2012tu}.

Given that in type IIB and F-theory only the K\"ahler moduli $T$ are {\it not}
stabilized by fluxes in a supersymmetric way, we now assume that supersymmetry
breaking is dominated by the $F$-terms of $T$.  However, our bulk Higgs from
D7-brane scalars has the peculiar feature that there is no K\"ahler moduli
dependence of the K\"ahler metric \cite{Aparicio:2008wh,Jockers:2004yj}. Hence,
assuming also shift symmetry and applying the supergravity formulae of
\cite{Brignole:1997dp}, the normalized soft Higgs masses are simply given by
\begin{equation}
m_1^2=m_2^2=m_3^2= 2m_{3/2}^2\,.
\end{equation}
The mass of the heavy Higgs doublet is therefore set to $m_S^2= 4 m_{3/2}^2\,.$

The Higgs mass matrix above is precisely of the form conjectured in a related
way in \cite{Hebecker:2012qp}. However, here it follows with much less model
dependence than expected earlier from a Giudice-Masiero-type analysis based on
higher-dimensional operators and $F_T$. In a sense, this is our main `success
story' concerning a framework with a natural stringy shift symmetry. The
remainder of this section is devoted to the related D7-brane intersection-curve
Higgs where, as it turns out, the picture concerning a shift symmetry (and a
possible prediction of the Higgs mass matrix in general) is much more
complicated. 

Most naively, the intersection-curve Higgs in type IIB orientifolds comes from
the intersection locus of a $U(1)$ brane and a $U(2)$ brane stack. The two
doublets $H_u$ and $H_d$ would in this case be two 4d zero modes of a 6d
hypermultiplet living on the intersection curve of the D7-brane-stacks.  This
has a straightforward generalization to the $SU(5)$ GUT case, where a ${\bf 5}$ and
${\bf \overline{5}}$ Higgs can originate in an analogous manner from a $U(1)$ and an
$SU(5)$ stack (or, even more generally, from the intersection locus of a $D_1$
and the GUT divisor). For such matter fields, it is known that the K\"ahler
potential has a K\"ahler-moduli-dependent prefactor \cite{Aparicio:2008wh}. We
will now describe a localization process turning the previously discussed bulk
Higgs into the intersection-curve Higgs. We aim in particular at understanding
what happens to the shift symmetry. 

To keep things as simple as possible, we ignore any possible
GUT-superstructure. Instead, we start from a $U(3)$ D7-brane stack and deform
it to the $U(1)\times U(2)$ stack. In this way, a `brane Higgs' at the
intersection curve can be continuously derived from a `bulk Higgs' contained in
the brane-deformation moduli $\zeta$ of the original $U(3)$ stack. 

The relevant K\"ahler potential is that of (\ref{kd7}), where we now assume
that $S$ and $z$ are stabilized by fluxes (we will suppress the $S$- and
$z$-dependence whenever possible) and Im$(S)\gg 1$:
\bea
\hspace*{-1cm}
K=-\ln[-i(S-\overline{S})-k_{D7}(z,\overline{z},\zeta+\overline{\zeta})]
&=& \frac{i}{(S-\overline{S})}\,k_{D7}(z,\overline{z},
\zeta+\overline{\zeta})+\cdots\nonumber \\ 
\\
&=&k_0\,\mbox{tr}(\zeta+\overline{\zeta})^2+\cdots\,. \nonumber
\eea
Here we have first Taylor-expanded in $k_{D7}$ (keeping only the linear term)
and then restricted ourselves to a quadratic approximation in $\zeta$ absorbing
the $S$- and $z$-dependence in the constant $k_0$. 

Our $\zeta$ is in the adjoint of $U(3)$, containing in particular two doublets
$H_u$ and $H_d$ in the complement of the $U(2)\times U(1)$ subgroup.  So far,
this is just group theory, but we go on to assume an actual deformation of our
brane stack such that the surviving gauge group is $U(2)\times U(1)$ and the
wave functions corresponding to the massless 4d fields $H_u$ and $H_d$ localize
along the emerging intersection curve
\cite{Conlon:2006tj,Beasley:2008dc,Font:2009gq}. The survival of $H_u$ and
$H_d$ as light 4d fields depends on geometrical data. More precisely, the
internal wavefunctions of localised massless modes correspond to elements of
the cohomology groups (see e.g. \cite{Beasley:2008dc} for a review)
\bea \label{Hi}
H^i(C, (L_a \otimes L_b^*)|_{C} \otimes K_C^{1/2}), \qquad i=0,1,
\eea
where $C$ is the intersection curve, $L_a$ and $L_b$ denote the line bundles of
the $U(2)$ and $U(1)$ branes whose curvature is the gauge flux $F_a$ and $F_b$
along the respective branes and $K_{C}$  represents the canonical bundle of
$C$. The modes corresponding to $i=0$ and $i=1$ characterize chiral ${\cal
N}=1$ multiplets in the representations $({\bf 2}_{1},-1)$ and $({\bf 2}_{-1},+1)$ under
$U(2)\times U(1)$. Note that the difference of the number of such fields is
given by the index $\int_C (F_a - F_b)$. In the sequel we assume that an
appropriate choice of fluxes is made corresponding to $h^i(C, (L_a \otimes
L_b^* )|_{C} \otimes K_C^{1/2}) =1$ for $i=0,1$. One possible configuration would be e.g. a setup
where $C$ is a $T^2$ and the fluxes are chosen such that $(L_a \otimes
L^*_b)|_C$ is the trivial bundle.

The relevant part of $K$ is\footnote{
In 
contrast to the rest of the paper, $H_u$ and $H_d$ are not canonically 
normalized in the present subsection.
}
\be
K \,\,\supset\,\, k_0 |H_u+{H}_d^\dagger|^2\,,
\ee
at least as long as the size of the $U(3)$-breaking deformation is sufficiently
small. By slight abuse of notation, we denote the size of this deformation by
$|\zeta|$. The `smallness' of this quantity is defined as the requirement that
the 8d wave functions of $H_{u,d}$ continue to be governed by the appropriate
section $H^0(D, N_{D/X})$ of the normal bundle of the original $U(3)$ divisor
$D$ inside the Calabi-Yau $X$,  such that the calculation of
\cite{Jockers:2004yj} continues to be valid. As already mentioned above, this
will cease to be correct for sufficiently large values $|\zeta|$, when the
relevant wave functions have become 6d wave functions described by (\ref{Hi})
and supported only on the intersection curve. 

In the case of torus orientifolds, the K\"ahler potential for D7-D7-brane
matter is known \cite{Aparicio:2008wh} (based on
\cite{Ibanez:1998rf,Lust:2004cx}).  In  particular, the kinetic term scales as
$1/\sqrt{st}$, where $s$ and $t$ are the real parts of $S$ and $T$. Thus,
focussing only on the $H_u$-kinetic term for brevity, we conjecture the
general, qualitative form
\be
K \supset \frac{k_0}{1+|\zeta|^2 \sqrt{t/s}} |H_u|^2 \sim \frac{1}{s+|\zeta|^2 \sqrt{ts}} |H_u|^2\,.
\label{qua}
\ee 
At $|\zeta|=0$, this is obviously in agreement with what we inferred from
\cite{Jockers:2004yj} before. At $|\zeta|^2\sim \sqrt{s/t}$, the typical
distance between the $U(2)$ stack and the $U(1)$ stack reaches string length
(one can convince oneself of this by returning to the original 10d string frame
action and carefully thinking about the definition of the relevant 4d
supergravity variables). Hence, at this value of $|\zeta|$, one expects a
transition to the new scaling regime advertised earlier. 

We give a more careful derivation of this transition in the Appendix. See also
\cite{Kawano:2011aa} for a discussion of the K\"ahler potential on matter curves. 

Qualitatively (though not up to ${\cal O}(1)$ factors) the above also applies
to the term $\sim |H_d|^2$. Unfortunately, we are unable to give a similar
argument for the terms $\sim H_u H_d$ and ${H}_d^\dagger {H}_u^\dagger$.  All
we can say is that the coefficients of these terms will also change dramatically
if $|\zeta|$ becomes so large that the Higgs fields localize on a curve.
Furthermore, this change will necessarily involve $s$ and $t$.  Given this
limited information, we parametrize the Higgs K\"ahler potential as
\be
K \,\,\supset\,\,f_1(T,\overline{T})|H_d|^2+f_2(T,\overline{T})|H_u|^2
    +f_3(T,\overline{T})H_u H_d + \mbox{h.c.}\,,
\label{fs}
\ee
and we know that, in the shift-symmetric limit, these three functions tend to a
single constant value, $k_0$ (cf. (\ref{qua})). Everything beyond that appears
to depend on the details of the geometry of the original $SU(3)$ surface and
the emerging intersection curve. Note that $f_1$ and $f_2$, are, at least in
principle, accessible by dimensional reduction at the two-derivative level.
This is not obvious for $f_3$ since the latter does not contribute to the Higgs
kinetic term before SUSY breaking. To approach the difficult issue of
determining the functions $f_i$ we next turn to a particularly simple setting.

\subsection{Toy model with a torus intersection curve and 
implications for more general situations}

Let us assume that our intersection curve is a $T^2$ and that the moduli space
of the Calabi-Yau orientifold with its branes has a locus where the induced
metric on this $T^2$ is flat. Furthermore, let us assume that within this locus
there is a region where our $T^2$ degenerates to an $S^1$. This very specific
simple case corresponds to the Higgs doublet pair being a hypermultiplet of a
5d gauge theory compactified on $S^1$. In this setting, and restricting
attention only to the modulus governing the radius of this $S^1$, concrete
statements can actually be found in the literature
\cite{Chacko:2000fn,Marti:2001iw}:

The 5d hypermultiplet possesses an $SU(2)_R$ symmetry which allows for a
`twisting' of the $S^1$-compactified theory by an element of this group.  This
is Scherck-Schwarz supersymmetry breaking and, moreover, it can be
characterized in 4d ${\cal N}=1$ language as spontaneous SUSY breaking
triggered by the $F$-term of the complexified $S^1$ volume modulus or radion
superfield. By slight abuse of notation, we call this superfield $T$ since it
morally corresponds to the type IIB K\"ahler moduli discussed before. The 5d
action can be displayed in 4d ${\cal N}=1$ notation\footnote{ 
Recall 
that, except in Subsection~\ref{bhvs}, we reserved the notation $H_u$, $H_d$ 
for canonically normalized Higgs superfields. Thus, obviously, field 
normalizations have to be adjusted: $H_u=\utilde{H}_u\sqrt{2\pi R(T\!+\!
\overline{T})}$ and analogously for $H_d$.
}
as \cite{Marti:2001iw}
\be
S=\int d^5x \left\{ \int d^4\theta (T+\overline{T}) (\utilde{H}_u 
\utilde{{H}}_u^\dagger + \utilde{H}_d^\dagger \utilde{{H}}_d)+\int d^2
\theta \utilde{H}_u \partial_5 \utilde{H}_d + 
\mbox{h.c.}\right\}\,.\label{s5d}
\ee
Crucially, no term $\sim \utilde{H}_u\utilde{{H}}_d$ appears.
Equivalently, as also worked out in \cite{Marti:2001iw}, the scalar mass matrix
induced by the SU(2)$_R$ twist is purely diagonal. 

Thus, in the extremely simplified setting discussed above, a shift-symmetric
structure does not arise. While an exchange symmetry appears natural 
(assuming that no gauge fluxes or other effects distinguish between the two 
complex scalars of our Higgs hypermultiplet), it can not be used to realize 
$\tan\beta=1$. The reason is that off-diagonal terms in the mass matrix are 
completely missing. In other words, tuning for a zero eigenvalue corresponds 
to tuning for a vanishing mass matrix so that $\tan\beta$ remains undetermined. 

This can be easily remedied by supplementing (\ref{s5d}) with a supersymmetric 
mass term $\mu$, which corresponds to the substitution $\partial_5\to 
\partial_5+\mu$ in (\ref{s5d}). One then obtains a Higgs mass matrix 
\cite{Marti:2001iw}
\be
m_1^2=m_2^2=|\mu|^2+\frac{|F_T|^2}{(T+\overline{T})^2}\qquad\mbox{and}
\qquad m_3^2=\frac{2\mu F_T}{T+\overline{T}}\,,\label{bmu}
\ee
where the required tuning for a vanishing determinant, $\mu=\overline{F}_T/
2\mbox{Re}T$, leads to $\tan\beta=1$ as desired. At the microscopic level, 
one can think of this $\mu$ term simply as of a 5d mass term or, more 
fundamentally, as of the vev of the $A_5$ or $A_6$ components in the 
underlying 5 or 6d gauged hypermultiplet theory.\footnote{
Recall 
that in 6d an ${\cal N}=2$ hypermultiplet can not have a mass term 
(see \cite{Fayet:1975yi,Strathdee:1986jr} and the discussion in Sect. 5 of 
\cite{Hebecker:2004xx}).
}

Alternatively, this mass term could remain zero at the classical level 
in the 5 or 6d theory, being generated by quantum effects in the 4d effective 
theory. For example, holomorphic $\mu$-terms can be induced by suitable 
D3-brane instantons along holomorphic divisors \cite{Blumenhagen:2006xt}, 
such that the size of the suppression of the coupling $\sim e^{\tau} H_u H_d$ 
depends on the specific regime in which the volume modulus $\tau$ of the 
instanton divisor is stabilized. This natural parametrical smallness may be 
advantageous. 

Returning to the classical 5 or 6d analysis, we note that generating the 
$\mu$ term through $A_5$ or $A_6$ expectation values corresponds to twisting 
the theory by a $U(1)$. Any diagonal $U(1)$ symmetry under which the scalars 
of the chiral components of the hypermultiplet transform in a vector-like way 
(i.e. $H_u$ and $H_d$ have opposite charges) is in principle suitable. For 
example, such an extra $U(1)\subset SU(2)_R'$ appears in trivial dimensional 
reduction of 10d SYM to 6 or 5d as a subgroup of transverse $SO(4)\sim 
SU(2)_R\times SU(2)_R'$ rotations. However, for $SU(2)_R\times SU(2)_R'$ to 
be a symmetry of a hypermultiplet, the latter has to be in a real 
representation of the gauge group (cf. \cite{Breitenlohner:1981sm} and 
Sect.~12.6 of \cite{Sohnius:1985qm}). In our context this presumably means 
that we would need to extend the Higgs field content. 

Let us now, based on the intuition gained from the oversimplified torus or 
even 5d case, return to a more generic setting. In other words, the two Higgs 
doublets now come from a hypermultiplet in a 6d model, compactified on some 
Riemann surface. In this case, the twisting is by $SU(2)_R$, by the $U(1)_T$ 
structure group of the tangent bundle, and by the gauge groups (including 
$U(1)$s) of the intersecting brane stacks. As long as the D7-branes are 
holomorphic divisors of a Calabi-Yau orientifold, the $SU(2)_R\times U(1)_T$ 
twisting leaves an ${\cal N}=1$ supersymmetry intact. By analogy to 5d radion 
mediation, one expects that SUSY-breaking by the $F$-terms of the K\"ahler 
moduli is equivalent to non-supersymmetric $SU(2)_R\times U(1)_T$ twisting. 
However, since the hypermultiplet scalars are singlets under the $U(1)_T$ 
structure group, this will leave us with an effective $SU(2)_R$ twisting in 
the scalar sector. This is similar to what we just saw in the 5d model of 
\cite{Marti:2001iw}. We are thus left without an off-diagonal mass term. As 
before, this can be cured by inducing a $\mu$ (and hence, as in (\ref{bmu}), 
a $B\mu$ term) by either non-perturbative effects or, classically, through 
twisting by gauge-theory $U(1)$s. This twisting might be associated 
with D7-brane Wilson lines or with D7-brane-fluxes. In the latter case, 
parametrical smallness of the $\mu$ term might be difficult to achieve. 

At a more fundamental level, the non-supersymmetric twists should 
be related to 3-form fluxes in the Calabi-Yau or, more precisely, to 
the warping induced by those fluxes \cite{Giddings:2001yu}. Thus, the exact 
form of the function $f_i(T,\overline{T})$ should be indirectly accessible 
through dimensional reduction in the presence of warping 
\cite{Burgess:2006mn}. In particular, it should be possible to understand 
in some detail how the Higgs mass matrix can be tuned by the flux choice. 
It would be very interesting but goes beyond the scope of this paper to 
actually perform such a calculation. 

To sum up, our best choice for finding a shift symmetry in a Higgs sector 
on D7-branes appears to be associated with a pure {\it bulk} Higgs. In this 
case, K\"ahler moduli do not enter the Higgs-sector K\"ahler potential and 
a simple mass matrix determined entirely by $m_{3/2}$ results. By 
contrast, looking for an exchange symmetry appears to be more promising in
situations with an {\it intersection-curve} Higgs. While the general analysis
is involved, the case with a $T^2$ intersection curve and no significant 
gauge flux effect seems to allow for the desired structure. The intermediate 
regime, where the Higgs is in transition from an 8d to a 6d field, is even 
harder to analyze and we were only able to give very qualitative results.

\section{${\cal N}=2$ $D$-term effects}
\label{sec:pterm}
Let us now pursue an alternative idea for realizing $\tan\beta=1$, based 
on $D$-term effects within the context of an ${\cal N}=2$ Super-Yang Mills 
theory. Our discussion is inspired by the `D6-branes-at-angles' mechanism 
for SUSY breaking with $\tan\beta=1$ of \cite{Ibanez:2012zg,Cremades:2002cs}.
We will attempt to formulate the central idea more generally and return to 
this particular scenario shortly. 

To begin, we recall that in a UV completion of the Standard Model any 
extended supersymmetry has to be broken at least at the compactification 
scale $m_C$ because of chirality.\footnote{
A 
model with an ${\cal N}=2$ breaking scale parametrically below $m_C$ is 
impossible since it would have to contain massive chiral fermions to pair 
up with those of the SM.
}
However certain non-chiral subsectors, such as the Higgs or the gauge sector,
may continue to enjoy an approximate ${\cal N}=2$ symmetry below $m_C$. By
contrast, it is also possible that breaking to ${\cal N}=0$ occurs directly at
the compactification scale. Thus, we see that a clear separation between the
various breaking scales from the higher-dimensional extended supersymmetry all
the way down to the non-supersymmetric SM can not be taken for granted. For
example in toroidal compactifications of type IIA (see
\cite{Blumenhagen:2006ci} and references therein), gauge theories on D6 branes
are locally $\mathcal{N}=4$, matter lives at points with $\mathcal{N}=1$ or
(real) curves with $\mathcal{N}=2$, and the overall supersymmetry of the
brane/o-plane network may be ${\cal N}=1$ \cite{Cvetic:2001nr,Honecker:2004kb} or even
$\mathcal N=0$ \cite{Blumenhagen:2000wh,Ibanez:2001nd,Blumenhagen:2001te}.
This
can have interesting implications for the Higgs mass matrix and quartic 
potential.

\subsection{Constraints on extended SUSY}
\label{sec:constraintsonextendedsusy}
Before coming to possible uses of extended SUSY, let us point out a crucial
constraint within our setting: We are interested in situations where
$(H_u,H_d^\dagger)$ form a hypermultiplet charged under an ${\cal N}=2$ U(1) gauge
theory. 
Recall that the $\mathcal N=2$ $D$-term potential \cite{Fayet:1975yi}
(also dubbed `$P$-term potential' \cite{Kallosh:2001tm}) arises from the terms 
\begin{equation}
{\cal L} \supset \cdots +\frac12 \vec{P}^2 + g\phi^A 
 \vec{P}\cdot\vec{\sigma}_A{ }^B  \,\phi^\dagger_B + \cdots
\end{equation}
of the complete lagrangian. Here $\vec{P}$ is the SU(2)$_R$-triplet auxiliary
field and $\{\phi^A\} \equiv (H_u,H_d^\dagger)$ are the scalars of the
hypermultiplet. The scalar potential hence reads 
\be
V_{D,\,{\cal N}=2}=\frac{1}{2}\vec{P}^2\,,     \label{vdn2}
\ee
where 
\be
P_3=g(|H_u|^2 - |H_d|^2)
\ee
and $P_1$, $P_2$ follow from $P_3$ by appropriate SU(2)$_R$ rotations of the 
SU(2)$_R$ doublet $(H_u,H_d^\dagger)$. From a 4D $\mathcal N=1$ perspective, 
$P_1+iP_2$ is an $F$-term while $P_3$ is a $D$-term. Thus,
\be 
V_{D,\,{\cal N}=2} = V_F + V_D = \frac{1}{2}|P_1+iP_2|^2 + \frac{1}{2}P_3^3\,.
\label{vdn22}
\ee
Explicitly, the potential takes the form
\begin{eqnarray}
V_{D,\,{\cal N}=2} &=& 
\frac12(-g \phi^A_\alpha \sigma^a{}_A{}^B\phi^{\dagger \alpha}_B)(-g \phi^C_\beta \sigma^a{}_C{}^D \phi^{\dagger\beta}_D)\nonumber \\
&=&\frac{g^2}{2}\Big((H_u H_d + H_d^\dagger H_u^\dagger)^2-(H_u H_d - H_d^\dagger H_u^\dagger)^2+(|H_u|^2 - |H_d|^2)^2\Big)\nonumber \\
&=& \frac{g^2}{2}\Big((|H_u|^2 - |H_d|^2)^2+ 4 |H_u H_d|^2\Big)\,.
\label{ourptermpotential}
\end{eqnarray}
Now we recall that our main interest in $\tan\beta =1$ was the implication 
$\lambda=0$, i.e., the flatness of the scalar potential for the 
light SM Higgs boson. This, however, is violated by $|P_1|^2+|P_2|^2$.
The $P_3^2$ term corresponds to 
the $D$-term potential 
\begin{equation}V= \frac{g^2}{2}(|H_u|^2 - |H_d|^2)^2\,,
\label{eq:standarddterm}
\end{equation}
which arises in this form since the 4D chiral multiplets $H_u$ and $H_d$ 
are in mutually complex conjugate representations of the gauge group. 

It is thus necessary to have all $D$- and $F$-terms which are not of the
type (\ref{eq:standarddterm}) decouple from the 
effective theory in an approximately supersymmetric fashion. This is 
tantamount to breaking to an MSSM-like gauge sector with $\mathcal N=1$ SUSY.
In other words, it appears to be necessary to have at least some small 
hierarchy between the scales where the gauge sector breaks to 
${\cal N}=1$ and to ${\cal N}=0$. Note that we talk here about the quartic
potential of the massless eigenstate. The potential (\ref{ourptermpotential}) alone
as a function of $H_u$ and $H_d$ has charge-violating flat directions of the
type $\langle H_u^0\rangle =-\langle H_d^-\rangle =c$, which are however not
preserved once the $SU(2)_L$-$D$-Term is included.

\subsection{Decoupling the `wrong' $D$-term potentials}

\begin{figure}
\begin{center}
\includegraphics[width=8cm]{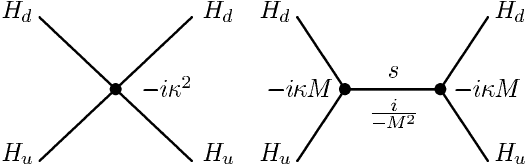}
\end{center}
\caption{Due to destructive interference between these diagrams, the quartic
$F$-term potential decouples at tree-level when the scalar $s$ is integrated out.
This decoupling is not exact if the scalar mass receives additional soft
breaking contributions.\label{integrateoutfterm}} 
\end{figure}

The mechanism of decoupling (and thus the appearance of flat directions in 
the low energy theory) can be understood from a simple 4D toy model describing
the extended SUSY sector in terms of superfields.
It includes a chiral singlet coupling to the Higgs doublets
\begin{equation}
\mathcal W = \kappa S H_u H_d + \frac12 M S^2 + \dots
\label{eq:ftermtoymodel}
\end{equation}
where $\kappa$ can be related to the gauge coupling in the previous section via $\kappa=\sqrt{2} g$.
The quartic Higgs coupling in such a simple model has for example been discussed 
in \cite{Giudice:2011cg}. In our case, we are especially interested in the 
interpretation $\sqrt{2}S=A_4+i\Phi_7$, appropriate to an ${\cal N}=4$ D6-brane 
theory with internal gauge components $A_4$, $A_5$, $A_6$ and adjoint scalars 
$\Phi_7$, $\Phi_8$, $\Phi_9$. We can then identify the chiral auxiliary field 
with two components of $\vec P$, e.g. $F^S \sim P_1+iP_2$. Above the scale 
$M$, the singlet and Higgs $F$-term contributions to the scalar potential are
\begin{eqnarray}
F^{S\dagger} F^S &=& \kappa^2 |H_u H_d|^2+M^2 S^\dagger S + 
\kappa M (H_u H_d S^\dagger + H_d^\dagger H_u^\dagger S) 
\label{eq:toymodelfterm}
\end{eqnarray}
and
\begin{eqnarray}
F^u F^{u\dagger} + F^{d\dagger}F^d &=& \kappa^2 S^\dagger S (|H_u|^2+|H_d|^2)\,.
\end{eqnarray}
This potential contributes to the quartic Higgs coupling and can modify the
tree-level prediction for $\lambda$. At the scale $M$, the entire chiral
multiplet $S$ is integrated out in an $\mathcal{N}=1$ supersymmetric fashion.
The corresponding Feynman graphs are shown in Figure \ref{integrateoutfterm}.
This can for example be the compactification scale. 
Since we expect the effective theory below $\Lambda=M$ to be a consistent 
supersymmetric theory at the renormalizable level, the $F$-term potential must 
vanish once the multiplet is integrated out. At tree-level, this happens via 
the exchange of the massive scalar in $S$ via the interaction in the third 
term of  (\ref{eq:toymodelfterm}). Indeed, when expanding the graphs in 
Figure \ref{integrateoutfterm}, one finds that the amplitude is of order 
$\mathcal O({p^2/M^2})$. There is therefore no dimension four quartic 
Higgs coupling in the effective theory\footnote{
This is of course a well-known property of SUSY theories, and essential e.g.
for SUSY GUT phenomenology. The same mechanism ensures that the divergent $F$- and
$D$-term contributions of Kaluza-Klein modes decouple in the low-energy limit of
extra dimensional models \cite{ftermdecoupling}.} 
below $\Lambda=M$. This decoupling of the $F_S$ potential is not exact any 
more once $s$ receives a soft mass $m_s^2$, and the potential resulting from 
the matching at treelevel to the theory without the scalar multiplet reads
\begin{equation}
V_{\Lambda=M} = \kappa^2  \frac{m_s^2}{m_s^2 + M^2} \, 
|H_u H_d|^2\stackrel{M\gg |m_s|}{\sim}\kappa^2  \frac{m_s^2}{M^2} \, |H_u H_d|^2
 \,.
\end{equation} 
As one would expect, the quartic coupling still vanishes for 
$M/m_s\rightarrow \infty$. However, for moderate hierarchies between the
overall soft scale and $M$, which could for example be the compactification 
scale (but also much lower), noticable corrections to $\lambda(m_S)$ and 
thus $m_h$ can appear. In models where no clear hierarchy between 
breaking to $\mathcal N=1$ and to $\mathcal N=0$ is present, this has to be 
taken into account. For example, if the SM $SU(2)$ gauge theory originates 
form a brane theory with extended SUSY, a direct breaking to ${\cal N}=0$ 
will lead to a significant quartic coupling $\sim g_2^2$ in the Higgs sector
{\it without} any flat direction.

\subsection{Using extended SUSY $D$-terms for the Higgs mass matrix}

Naively, one would conclude from the above that all gauge theories under 
which our hyper is charged have to be broken to the `surviving' ${\cal N}=1$ 
SUSY much above the `last' SUSY breaking scale. In this case, the quartic 
potential is flat along the direction of the SM Higgs if $\tan\beta=1$ can be 
realized. 

If, at the same time, we want to use a non-zero (field-dependent) FI term 
of such a gauge theory to generate a contribution to the Higgs mass matrix,
\be
{\cal L}\supset g^2 \left( \xi + |H_u|^2-|H_d|^2\right)^2\,,  \label{fi}
\ee
one encounters a problem: One finds opposite contributions to the soft 
masses of $H_u$ and $H_d$, ruining $\tan\beta=1$. Thus, $\xi=0$ appears to 
be required and no useful effect of our extended-SUSY $D$-terms is left. 

However, since we are here {\it not} dealing with a SM gauge group, we can 
try to avoid such a pessimistic conclusion by considering very small $g$. 
(Formally, this would be the limit $g\to 0$ with $g^2\xi$ fixed.) 
In this case, the induced quartic potential vanishes but the mass correction 
from $\xi$ survives. The vanishing of the quartic potential allows us to 
consider an ${\cal N}=2$ $U(1)$ theory, which in turn gives us the choice 
between three FI terms. They are related by an $SU(2)_R$ rotation, which of 
course also affects our Higgs scalars. An interesting choice is to set the 
`usual' FI term (the one associated to $P_3$) to a non-zero value, but 
working in a frame in which the hyper has the form 
\be
\{\Phi^1,\Phi^2\} \equiv \{ (H_u-H_d^\dagger) , (H_d^\dagger+H_u) \}\,.
\label{rotatedhyper}
\ee
Then, in analogy to (\ref{fi}), we find
\be
{\cal L}\,\,\supset\,\, g^2 \left( \xi + |H_u-H_d^\dagger|^2-|H_d^\dagger+
H_u|^2\right)^2 \,\,\supset\,\, -4 g^2\xi H_u H_d + \mbox{h.c.}\,,
\label{fir}
\ee
which gives us a $B\mu$ term from a $D$-term. This is our interpretation of 
the very interesting suggestion for realizing the exchange symmetry (and 
hence $\tan\beta=1$) in the context of intersecting D6 branes in 
\cite{Ibanez:2012zg}. Let us now scrutinize this particular implementation 
of the field-theoretic mechanism described above in more detail:

We recall that, in this context, the Higgs hypermultiplet lives at the
intersection of a $U(2)$ stack of D6 branes (D6$_a$) with a single D6 brane
(D6$_b$), where for definiteness we are working in the context of Type IIA
orientifolds with D6-branes instead of their T-dual Type IIB version. While we
assume the $SU(2)$ inside $U(2)$ to be the group of SM weak interactions, we
remain agnostic concerning the SM hypercharge $U(1)_Y$. The latter can be some
linear combination of the two $U(1)$s just introduced (or even just the $U(1)$
from D6$_a$) and further $U(1)$ groups. 

To obtain a hypermultiplet Higgs, D6$_a$ and D6$_b$ have to intersect 
(non-generically) in a real line rather than (generically) in a point. 
The role of the ${\cal N}=2$ $U(1)$ theory can then be played by the gauge 
theory on D6$_b$, with $\xi$ related in the standard way to a small deviation 
of the intersection angles from the SUSY condition. The smallness of $g$ 
corresponds to a large volume of D6$_b$. 
\begin{figure}[h]
\begin{center}
\includegraphics[width=14cm]{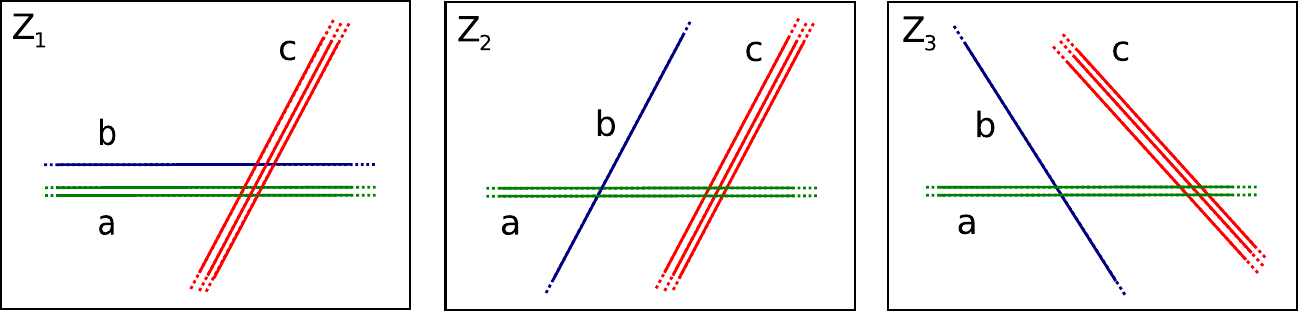}
\end{center}
\caption{Sketch of the local intersection structure of the three relevant 
brane stacks in the D6-brane-realization of an exchange-symmetric Higgs mass
matrix.\label{stacks}}
\end{figure}

We now want to break the supersymmetry of the SM $SU(2)$ brane stack D6$_a$
in the appropriate way. Before doing so, let us first recall the situation 
in a standard ${\cal N}=1$ compactification with ${\cal N}=2$ Higgs sector.
Without loss of generality, we start with a brane stack D6$_a$ filling the 
$468$ plane, and choose a complex structure by defining that
\begin{equation}(z_1,z_2,z_3)=(x^4+ix^5,x^6+ix^7,x^8+i x^9)\,.
\label{eq:complexstructure}
\end{equation}
In flat space, we can think of $z_i$ as transforming like a fundamental 
$\bf 3$ of the ``Calabi-Yau like'' holonomy group $SU(3)\subset
SO(6)$.  As long as we remain in the standard $\mathcal N=1$ picture, all D6/O6
positions are related to D6$_a$ by such $SU(3)$ rotations. For example, we can
obtain D6$_b$ from D6$_a$ via 
\begin{equation}
T(M_{b})= {\rm diag}(1,e^{i\theta},e^{-i\theta})\in SU(2)\subset SU(3).
\end{equation}
Here, we denote by $M_i$ the $SO(6)\subset SO(1,9)$ rotations in the spinor representation
while $T(M_i)$ is the corresponding representation of $U(3)\subset SO(6)$ acting on (\ref{eq:complexstructure}).
For our purposes, it is sufficient to supplement D6$_a$ and D6$_b$ with 
a (SM color) $SU(3)_c$ stack\footnote{ Actually,
the $SU(3)$ stack is introduced just as a simple illustration of the 
required SUSY breaking to ${\cal N}=1$ of D6$_a$. What is crucial for 
all that follows is a reduction of the spectrum on D6$_a$ to that
of an ${\cal N}=1$ gauge theory. This is not realized by the 
$SU(3)$ stack, but by intersection with an O-plane. We assume that this latter
intersection breaks SUSY in the same way as the $SU(3)$ stack and keep 
talking about the $SU(3)$ stack as a simple way of specifying a
particular ${\cal N}=1$ subalgebra. 
} D6$_c$, which has a chiral intersection with 
D6$_a$, as shown in Fig.~\ref{stacks}. It is obtained from D6$_a$ via a rotation 
\begin{equation}
T(M_c)= {\rm diag}(e^{i\theta_1},e^{i\theta_2},e^{i\theta_3})
\end{equation}
with $\theta_1+\theta_2+\theta_3=0\mbox{ mod }2\pi$ as usual.
As a result, the D6$_{a,b}$ system enjoys ${\cal N}=2$, while the D6$_{a,c}$
system only enjoys ${\cal N}=1$ supersymmetry.\footnote{For a general
discussion of the relative supersymmetry preserved by branes at angles we refer
to \cite{Berkooz:1996km}.}

The position of D6$_b$ is characterized by two angles $\theta_2$ and $\theta_3$
in the $z_2$- and $z_3$-plane. As long as $\theta_2+\theta_3=0$, this is just a
parametrization of the $SU(2)$ rotation introduced above.  We now go beyond
$SU(2)$ to $U(2)\not\subset SU(3)$ by allowing for a small non-zero
$\delta\equiv \theta_2+\theta_3$, characterizing supersymmetry breaking from
${\cal N}=2$ to ${\cal N}=0$. It is well known\cite{Cremades:2002cs,Ibanez:2001nd} that, if we separate D6$_a$ and
D6$_b$ along Im$(z_1)$ to avoid a tachyonic instability, then the mass matrix
for the fields $\{\Phi^1,\Phi^2\}$ at the real intersection curve takes the
form\footnote{Other effects of D-term breaking on the spectrum in models with extended relative supersymmetries have been discussed recently in \cite{Anastasopoulos:2011kr}.} 
\be 
\left( \begin{array}{cc}
\mu^2 + \alpha \delta &       0               \\
      0               & \mu^2 - \alpha \delta
\end{array} \right)\,.                        \label{mats}
\ee
We would, by contrast, like to obtain the form
\be 
\left( \begin{array}{cc}
     \mu^2            & \alpha \delta \\
\alpha \delta         &      \mu^2            \label{matb}
\end{array} \right)\,,
\ee
which corresponds to a rotation of (\ref{mats}) by 
\be
U = \frac{1}{\sqrt{2}}
\left( \begin{array}{cc}
 \,1 & \!1 \\
-1 &  1
\end{array} \right)\in SU(2)_R\,.
\ee

This last rotation is automatically realized if we manage to ensure that 
$\{\Phi^1,\Phi^2\}=\{ (H_u-H_d^\dagger) , (H_d^\dagger+H_u) \}/\sqrt{2}$, with $H_u$, $H_d$ 
being the scalars of the superfields corresponding to the ${\cal N}=1$ SUSY
respected by the D6$_{a,c}$ system. Thus, we need to `rotate' the SUSY breaking 
induced by D6$_c$ on the D6$_a$ gauge theory by an $SU(2)_R$ twist $U$. 
 All we would then have to do is to replace D6$_c$ by D6$_c'$, where the position of 
the latter follows from D6$_a$ via a rotation by $U$ 
\be
M_c'\equiv UM_c U^\dagger\,,
\ee
where all objects are now matrices acting on the spinor representation of
$SO(1,9)$.  What is the appropriate choice for $U$?  The condition for the
unbroken supersymmetry parameter in the $D6_{a,c}$ system is
\begin{equation}
[\Gamma_{D_a},M_c]\epsilon=0,
\end{equation}
where $\Gamma_{D_a}=\prod_i e_i^M \Gamma_M$ is defined in the usual
fashion\cite{Berkooz:1996km}, and $i$ runs over the brane-parallel directions.  
We want to implement $U$ such that the new condition for the unbroken
parameter $\epsilon'$ of the $D6_{a,c'}$ system,
\begin{equation}
[\Gamma_{D_a},M'_c]\epsilon'=0,
\end{equation}
automatically holds for $\epsilon' = U\epsilon$. Furthermore, $\epsilon'$
should precisely correspond to the $\mathcal N=1$ supersymmetry in which the
hypermultiplet takes the desired form (\ref{rotatedhyper}).  The position of
the D6$_a$ stack in the $468$ direction as depicted in Figure \ref{stacks}
corresponds to 
\begin{equation}
\Gamma_{D_a}=\Gamma_4\Gamma_6\Gamma_8\,.
\end{equation}
The real line in the $D_{a,b}$ system on which the hypermultiplet resides was
obtained by setting $\theta_1=0$. This determines the two supersymmetries of
the Higgs system.  To construct $U$ in the spinor representation, we now
consider the $SO(4)\subset SO(6)$ which rotates the coordinates $6789$ while
leaving the real line of the Higgs system untouched.   The generators of this
$SO(4)\cong SU(2)\times SU(2)_R$ can be represented as
\begin{eqnarray}
T^{\phantom R}_1=\frac{i}{8}\big([\Gamma^6,\Gamma^9]-[\Gamma^7,\Gamma^8]\big)\,,
T^{\phantom R}_2=\frac{i}{8}\big([\Gamma^6,\Gamma^8]-[\Gamma^9,\Gamma^7]\big)\,,
T^{\phantom R}_3=\frac{i}{8}\big([\Gamma^6,\Gamma^7]-[\Gamma^8,\Gamma^9]\big)\,
\nonumber \\
T^R_1=\frac{i}{8}\big([\Gamma^9,\Gamma^6]+[\Gamma^8,\Gamma^7]\big)\,,
T^R_2=\frac{i}{8}\big([\Gamma^6,\Gamma^8]+[\Gamma^9,\Gamma^7]\big)\,,
T^R_3=\frac{i}{8}\big([\Gamma^6,\Gamma^7]+[\Gamma^8,\Gamma^9]\big).\nonumber\\
\end{eqnarray}
The subgroup $SU(2)$ is contained in the $SU(3)$ ``holonomy'' group and leaves
the unbroken supersymmetry invariant.  The subgroup $SU(2)_R$ however does not
respect the complex structure, and we can use it to rotate the unbroken
supersymmetry (note however that arbitrary $SU(2)_R$ transformations on $M_c$
can break supersymmetry in the D6$_{a,c'}$ system completely). The generator
$T^R_3$ performs a ``factorizable'' rotation and therefore commutes with $M_c$.
It leaves $M_c$, and consequently the unbroken supersymmetry, invariant.  The
generator $T^R_2$ on the other hand satisfies $[T^R_2,\Gamma_{Da}]=0$ since it
leaves the branes in D6$_a$ manifestly invariant.  As a consequence, for the
choice 
\begin{equation}
U=\exp[{i \frac{\pi}{2}\, T^R_2}]
\end{equation} 
the SUSY condition for the D6$_{a,c'}$ system becomes
\begin{equation}
0\stackrel{!}{=}[\Gamma_{Da},M_c']\epsilon'=[\Gamma_{Da}, U M_c U^\dagger]\epsilon'=U[\Gamma_{Da},M_c]U^\dagger \epsilon'
\end{equation}
which is satisfied for $\epsilon' = U \epsilon$.   
$U$ now rotates
the two unbroken supersymmetry parameters of the Higgs sector into each other
(by an angle of $\pi/2$).  As a result, the ${\cal N}=1$ supersymmetry of D6$_a$
respected by the intersection with D6$_c'$ is precisely the one in which the
${\cal N}=1$ superfields $H_u$ and $H_d$ receive a $B\mu$ term mass correction
$\sim \delta$ according to (\ref{fir}). 

A different perspective might also be useful: Let us stick with $M$ as the
rotation of D6$_a$ to D6$_c$ and denote by $(1+\delta)N$ the rotation of 
D6$_a$ to D6$_b$. Here $\delta$ is now a small matrix, such that $N\in SU(2)$ 
and $(1+\delta)\in U(2)$. Rather than rotating $M$ by $U$, we instead rotate 
$(1+\delta)N$ by $U^\dagger$, which is clearly equivalent. We thus have
\be
(1+\delta)N \to U^\dagger(1+\delta)NU = (1+U^\dagger\,\delta\, U)N\,.
\ee
The last equality follows since $N\in SU(2)$ and $U\in SU(2)_R$, with
$SU(2)$ and $SU(2)_R$ the two commuting factors of $SO(4)$ or, equivalently, 
the upper-left and lower-right $2\times 2$ blocks of $SU(4)$. 
We now see very clearly that one can modify the SUSY-breaking rotation 
$(1+\delta)$ such that it induces an $F$-term instead of a $D$-term. This
is not what is usually done in SUSY-breaking by angles, since one usually 
just modifies angles in the factorized geometry of three complex planes. 
But this is clearly not the generic possibility of rotating branes, and the
appearance of an $F$-term corresponds to a general rotation after which the
D6-brane ceases to be a lagrangian brane.  In a Type IIB dual such an $F$-term
translates into $F$-term breaking by non-holomorphicity of the branes and/or by
suitable gauge fluxes. Note that the latter option requires gauge fluxes which
are not the pullback of ambient space forms and is thus not available in simple
toroidal geometries.

In summary, we have seen that deriving a $B\mu$ term from the $D$-terms of 
extended supersymmetry is field theoretically possible, but a detailed 
understanding of the hierarchies of SUSY breaking is mandatory. In particular,
in order to realize an exchange-symmetric Higgs mass matrix, one has to make 
sure that the soft diagonal contributions are strongly suppressed.
For example, any model that exploits $\tan\beta=1$ to predict approximately 
vanishing quartic coupling but generates $B\mu$ via $F$-term contributions
must ensure that the same $F$-term decouples sufficiently from the Higgs 
potential below $m_S$. This might be possible in the limit where the 
effective gauge coupling of the gauge theory from which this $F$-term 
originates is small. Finding an explicit stringy realization with D6 branes on 
tori appears doable according to the discussion given above. 
We stress, however, that the torus geometry, which is crucial for the viability
of the SU(2)$_R$ symmetry underlying the argument, is at odds with a strong
hierarchy between the gauge coupling associated with the theory from the which
the $F$-term results and the Standard Model: Due to the simple cycle structure on
$T^6$ the only way to ensure strongly differing gauge couplings and thus brane
volumes is to include branes with high wrapping numbers, which, on the other
hand, is constrained by tadpole cancellation conditions.  More interesting 
would be an understanding of this mechanism in the context of a proper 
Calabi-Yau geometry.

\section{Corrections to the Quartic Coupling}
\label{sec:corrections}
\subsection{Radiative Corrections to the Quartic Coupling}
\label{sec:radcor}
The dominant radiative corrections to the quartic coupling in the UV come from
two sources: shift/exchange symmetry violating corrections to the Higgs mass
matrix will lead to a deviation from $\tan\beta=1$ and thus indirectly to
corrections of the effective quartic coupling in the SM. Furthermore, the
quartic coupling itself receives threshold and decoupling contributions when
the heavy Higgs states, sfermions (in particular the top partners) and
higgsinos and gauginos are integrated out near $m_S$. In unbroken SUSY, the
relation (\ref{lambda}) between the $D$-term quartic coupling and the gauge
couplings generally only holds to higher orders in manifestly supersymmetric
renormalization schemes like $\overline{DR}$.  We find that the corresponding
conversion of the quartic coupling to $\overline{MS}$ yields a numerically
negligible contribution for our purposes.
Our discussion of radiative corrections to the
Higgs mass in high-scale supersymmetry is somewhat analogous to the one
performed in \cite{Hall:2009nd}. However, while these authors have concentrated
on the case $\cos^2{2\beta}\sim 1$ which is now disfavored by data, we treat
the complementary case $\tan\beta\sim1$.

\paragraph{Corrections to the Higgs Mass Matrix:} 
In a previous paper \cite{Hebecker:2012qp}, we have studied the
first effect in order to quantify the impact of shift symmetry violating
interactions on the observed Higgs mass. The starting point is the usual mass
matrix with universal entries (\ref{mi})
from which one obtains $\tan\beta=1$ and thus $\lambda(m_S)=0$ from the tree
level $D$-term potential (see Section \ref{sec:tanbeta}).  Radiative corrections,
in particular from the superpotential Yukawa couplings to the top quark,
necessarily modify this texture at 1-loop, leading to a deviation from $\cos^2
2\beta=0$. For shift symmetric scenarios in which the symmetry can be assumed
to be exact in the uncompactified higher dimensional theory, there are two
related corrections to the mass matrix. Since the transition from the MSSM to
the SM is not at the compactification scale $m_C$, but rather at a lower scale
$m_S$, we can collect the effects from the shift symmetry violating
interactions by the RGE running from $m_C$ down to $m_S $\footnote{The same
strategy can be applied to estimate the corrections to the exchange symmetric
scenarios. In \cite{Ibanez:2012zg}, the authors estimate the corrections to the
Higgs mass with a different approach, but essentially arrive at the same
conclusions}.
The corrections are of the form\cite{Hebecker:2012qp}
  \begin{equation}\frac{\delta m_{Hi}^2}{ m_{Hi}^2}\sim \,\frac{6 \overline{y_t^2}}{16 \pi^2}\log\left(\frac{m_S}{m_C}\right)\end{equation}
and consequently, the shift violating (SV) correction to the quartic coupling is given by
\begin{equation}\delta \lambda_{SV}(m_S)\sim C\,\frac{g_2^2+g_1^2}{8}\,\Big|\frac{6 \overline{y_t^2}}{16 \pi^2}\log\left(\frac{m_S}{m_C}\right)\Big|^2\,.
\label{shiftdeltalambda}
\end{equation}
where the constant $C$ depends on how the electroweak scale is tuned light and
was assumed to be $C=2$ for definiteness.  When integrating out $\tilde t_i$,
we get decoupling contributions to the squared Higgs masses from quadratically
divergent self energy graphs. At 1-loop, they include an ambiguity concerning
the choice of matching scale and amount to an  additive constant of
$\mathcal{O}(1)$ to the log. For $m_C\gg m_S$, the constant is therefore
negligible compared to the log and can be absorbed in a slight shift of $m_S$.

\paragraph{Integrating out heavy MSSM states:}
The second type of radiative corrections originating from the stops (T) is
the one to the quartic coupling itself\cite{Okada:1990vk},
\begin{equation}\delta \lambda_{T}(m_S) = \frac{3 y_t^4}{16 \pi^2}\Big[
\frac{X_t^2}{m_{\tilde t}^2}\Big(1-\frac{X_t^2}{12 m_{\tilde
t}^2}\Big)+2\log(\frac{m_{\tilde t}}{m_S})\Big]
\label{lambdathreshold}
\end{equation}
where $X_t=A_t - \mu \cot \beta \approx A_t - \mu$ is the effective trilinear
coupling. We have assumed negligible splitting among the top partners.  The correction
$\delta \lambda_T$ is
nominally suppressed by one loop factor less than $\lambda_{SV}$ in Eq.
(\ref{shiftdeltalambda}).  The first term\footnote{We thank Manuel Drees for
reminding us of the potential importance and different quality of this contribution.}{$^,$}\footnote{Note that a negative correction to $\lambda$ can be obtained from the 1-loop stop decoupling corrections 
for $X_t > \sqrt{12} m_{\tilde t}$. A nonperturbative regime can arise from extreme values of $X_t$, and
one has to be careful not to introduce charge- or color breaking 
minima\cite{Cornwall:2012ea}.} in the brackets
in Eq. (\ref{lambdathreshold}) is the 1-loop effective $A$-term
contribution from the finite graphs with four external SM Higgs fields.  The
second term encodes the threshold which arises if the stops are not
integrated out at $m_{\tilde t}$.
While the SM running of the quartic coupling at high scales becomes increasingly
flat due to a cancellation between Yukawa and gauge contributions, and the
overall sensitivity on $m_S$ is thus suppressed, this is not necessarily true
for the individual thresholds. Thus, in addition to the effects discussed in
\cite{Ibanez:2013gf}, we found that for moderate splittings $m_{\tilde t}/m_S$  
and intermediate values of $X_t$, the effective $A$-term correction and thresholds
in (\ref{lambdathreshold}) are of similar size and can cancel. 

\begin{figure}
\begin{center}
\begin{picture}(500,160)
\put(0,50){\rotatebox{90}{$m_h/GeV$}}
\put(180,-10){$\log_{10}(m_S/GeV)$}
\put(15,10){
\includegraphics[width=7cm]{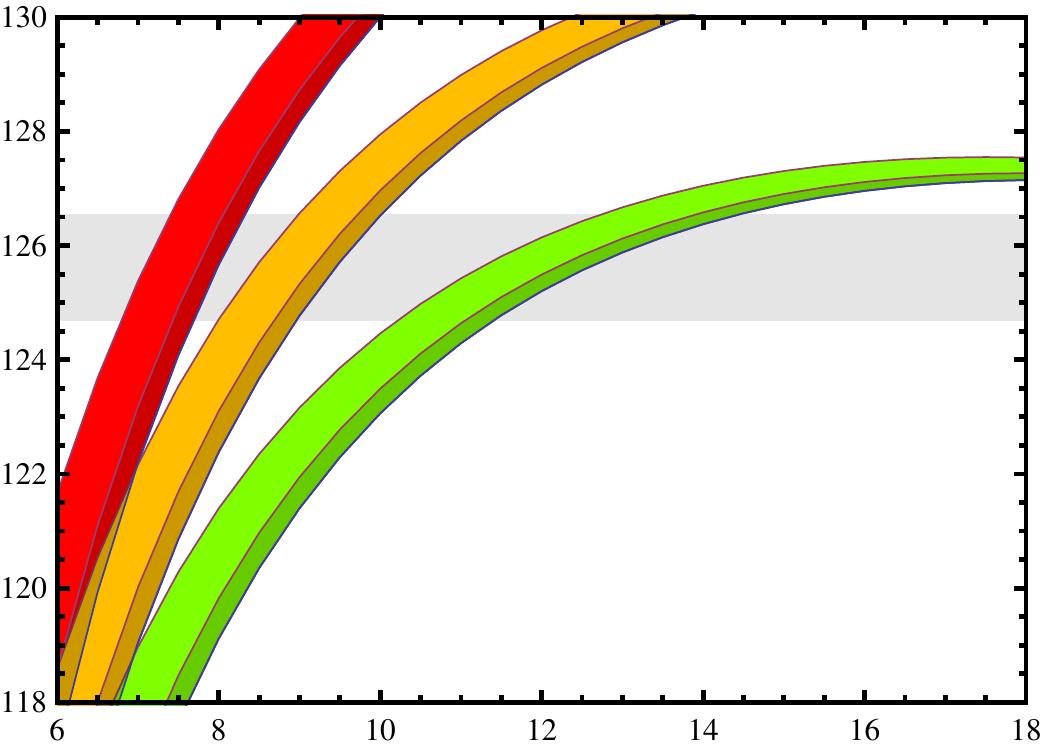}\qquad
\includegraphics[width=7cm]{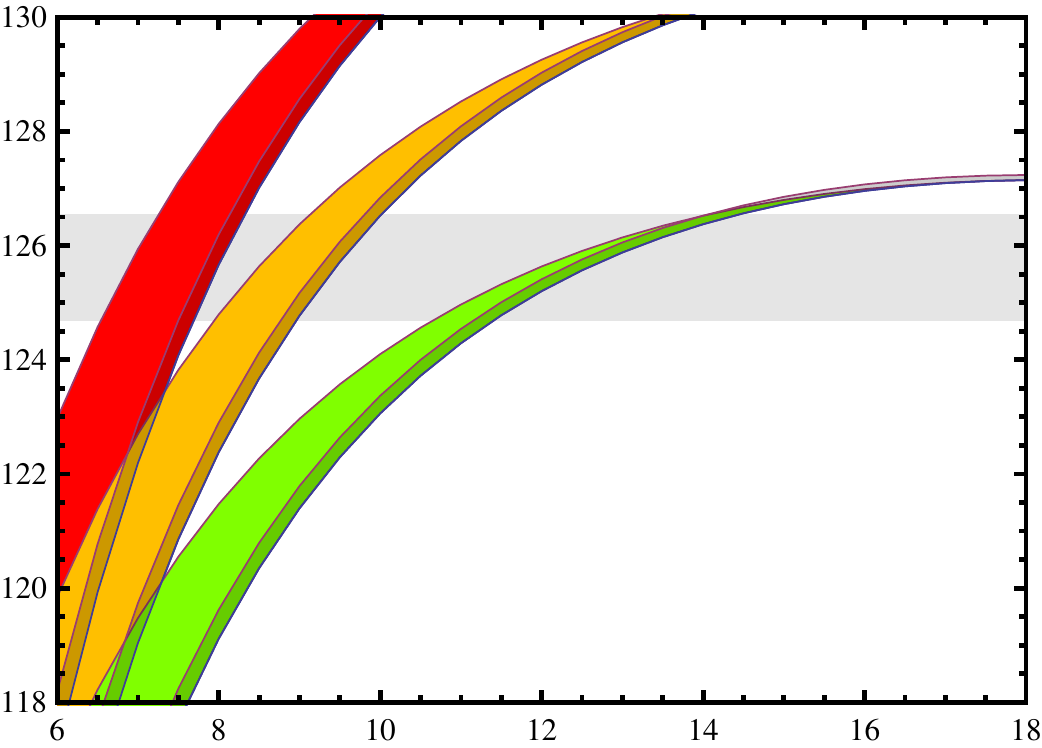}}
\end{picture}
\end{center}
\caption{The impact of squark decoupling corrections to the quartic Higgs
coupling (left) and shift/exchange symmetry violation (right) on the physical
Higgs mass.  The narrow dark(broad light) bands are for $X_t^2=m_S^2\,(6
m_S^2)$ for the decoupling contributions from top partners, and $m_C=10^2\, m_S
(\sqrt{m_S m_{Pl}})$ for the shift symmetry violation. The top quark masses are
$m_t=175.5,173.5,171.5$ from upper (red) to lower (green) band. The scale $m_S$
should be understood as the effective SUSY scale as defined in
(\ref{eq:effectivesusyscale}).\label{fig_thresholdhiggsmasses}}
\end{figure}

Assuming that $X_t$ and $m_{\tilde t}$ are roughly of the same order (splitting
between $\tilde t_{L,R}$ is neglected here), the first
contribution is maximized for $X_t^2=6 m_{\tilde t}^2$. For $X_t^2=0\dots 6
m_{\tilde t}^2$ it is in the range
\begin{equation}
\delta \lambda_{T}(m_S) = 0\dots 3\times \frac{3 y_t^4}{16
\pi^2}\,.
\end{equation} 
The calculation given in \cite{Hebecker:2012qp}, where the focus was on effects
from shift symmetry violation and the viability of the shift symmetric
scenario, does not include these corrections. However, they should be taken
into account together with the SUSY thresholds and the shift violating
contributions to get a meaningful estimate of the value and uncertainty in
$m_S$.  The impact of these radiative corrections to $\lambda$ on the physical
Higgs mass is illustrated in Figure \ref{fig_thresholdhiggsmasses} for
$\delta\lambda_{SV}$ and $\delta\lambda_{T}$.  The shaded Higgs mass range
corresponds to the union of the ATLAS and CMS (1$\sigma$) errors 
quoted with the observation \cite{expatlas,expcms}, with
statistics and systematics added in quadrature. The three colored bands
correspond to the current central value given by PDG for the top quark mass
($173.5\pm.6\pm.8$ GeV,\cite{pdgtop}) and $\pm 2\sigma$, again with errors added in
quadrature.  It is noteworthy that both the shift symmetry violating
corrections and the threshold corrections to $m_h$ owe their relative smallness
to the RGE evolution of $y_t$ towards smaller values in the UV with
$y_t^4(10^9,\,10^{16},\,10^{19} \mbox{ GeV}) \approx 1/9,\,1/27 ,\,1/40$.

Let us now consider the remaining threshold corrections in the limit
$\tan\beta=1$. They can be found e.g. in
\cite{Giudice:2011cg,Dobado:2002jz}.  The higgsino and gaugino threshold
contributions are somewhat more complicated than the stop thresholds. We take
the expressions given in \cite{Giudice:2011cg} and apply some simplifications.
First of all, we assume 
\begin{equation}
r_1\equiv\frac{M_1}{\mu}=r_2 \equiv \frac{M_2}{\mu}=\frac{M_\lambda}{\mu}\equiv r,\quad \tan\beta=1,\quad \lambda^{LO}=0\,,
\end{equation}
where the higgsino mass $\mu$ should not be confused with the renormalization
scale, and $M_1,M_2$ are the electroweak gaugino masses.  The corrections to
$\lambda(m_S)$ given in \cite{Giudice:2011cg} then reduce to
\begin{equation}
\delta\lambda_{GH}(m_S)=\frac{\tilde b_\lambda}{16 \pi^2}\,\left [\log \frac{\mu}{m_S} + \frac{(r-1)(r+1)^2 +2(r-3) r^2 \log r}{2 (r-1)^3}\right]
\end{equation}
where
\begin{equation}
\tilde b_\lambda= \frac12(-g_1^4-2 g_1^2 g_2^2 -3 g_2^4)\,.
\end{equation}
In leading log approximation this becomes
\begin{equation}
\delta \lambda_{GH}\approx \frac{\tilde b_\lambda}{16 \pi^2} \log \frac{m_\chi}{m_S}
\end{equation}
where $m_\chi \equiv \max(\mu,M_\lambda)$.
Note that we define $g_1$ in SM normalization rather than GUT normalization,
and that our quartic coupling is normalized as $2\lambda(\mbox{this
paper})=\lambda(\mbox{\!\!\cite{Giudice:2011cg}})$.

Finally, there is a threshold from the heavy Higgs states, which will complete
the MSSM-like 4D spectrum. We use the result in \cite{Dobado:2002jz}, again in
the limit $\tan\beta=1$, and find 
\begin{equation} \delta
\lambda_{A}=-\frac{1}{16 \pi^2} \frac14 \tilde b_\lambda \log\frac{m_A}{m_S}
\end{equation}
where $m_A$ is the mass of the heavy Higgs doublet, in our case $m_A^2 \sim 2
B\mu$. The constant decoupling contributions from mixed heavy-light Higgs
diagrams are found to vanish in the limit $\tan\beta=1$. As a quick
consistency check\footnote{We find a discrepancy between the heavy Higgs thresholds
given in \cite{Dobado:2002jz} which we use here, and \cite{Giudice:2011cg}.},
note that the logs in $\delta \lambda_{T}+\delta
\lambda_{GH}+\delta\lambda_{A}$ reproduce the leading log of the SM running of
the quartic coupling. Specifically, in the
$\lambda^{LO}=0,\tan\beta=1$ limit, we obtain 
\begin{equation}
16 \pi^2 \frac{\partial\,}{\partial \log m_S}\delta\lambda_{T+GH+A} = \left[-\frac34 \tilde b_\lambda - 6 y_t^4 \right]=b_\lambda^{SM,1-loop}\Big|_{\lambda=0}\,.
\end{equation}
which means that the quartic coupling obtained from the SUSY theory as a function of $m_S$ ``runs'' 
like the SM quartic coupling to leading log precision. This lets the unphysical matching scale $m_S$ drop out
of our Higgs mass prediction (for small variations of $m_S$). 

Ignoring the constant thresholds for now, one can find the value of the
unphysical scale $m_S$ for which the threshold logs cancel among each other. 
This ``effective SUSY scale'' to leading log precision, $m_S=m_S^{eff}$, 
is given by 
\begin{equation}
\label{eq:effectivesusyscale}
m_S^{eff}=\left[m_A^{-\tilde b_\lambda/3} m_{\tilde t}^{8 y_t^4}m_\chi^{4 \tilde b_\lambda/3}\right]^{1/(\tilde b_\lambda + 8 y_t^4)}\,.
\end{equation}
Thus, in leading log approximation, this scale choice allows us to set the
quartic coupling to the tree level relation\footnote{Since we have used the
approximation $\cos^2 2\beta=0,\lambda^{LO}=0$ in the threshold formulae, this
is only consistent if $\cos^2 2\beta$ is of loop-suppressed size. This is
generally the case in the shift/exchange symmetric models which we discuss in
this paper.} $\lambda(m_S^{eff})=(g_1^2(m_S^{eff})+g_2^2(m_S^{eff}))\cos^2{2\beta}/8$. 
The denominator of the exponent in Eq. (\ref{eq:effectivesusyscale}) corresponds
to the SM $\beta$ function of $\lambda$. For
vanishing SM running, $m_S$ is therefore undetermined in this approximation.
Note that in our derivation of the scale $m_S^{eff}$, we have assumed that the 
constant thresholds can be neglected.

What is the relative importance of the threshold contributions compared to the
effective $A$-term contribution? It turns out that the former can be negative and
of similar magnitude as the latter even for the extreme case $X_t^2=6 m_{\tilde
t}^2$. For example, for $m_S=10^{9}$ GeV, we have $y_t^4\sim 0.11$ and $\tilde
b_\lambda\sim -0.24$. We choose the matching scale at the mass scale of the
heavy Higgs doublet, $m_S^2= m_A^2\sim2 B\mu$, thus resumming the log in
$\delta \lambda_{A}$. For definiteness, consider a scenario where the Higgs
mass matrix has no soft mass contributions. This situation occurs in some
models discussed in this paper.  Then, $\mu^2\sim B\mu$, and thus $ \mu \sim
m_S/\sqrt{2}$. For vanishing splittings $r=1$ and $m_{\tilde t}\sim \mu$, we
get $\delta\lambda_{GH}\sim \tilde b_\lambda/(4\pi)^2 \sim -0.0015$ and
$\delta\lambda_{T, log}\sim-0.0015$. Although we have $\tilde
b_\lambda\log(\mu/m_S)>0$, the total GH contribution is negative because the
constant part dominates over the log for $r\rightarrow 1$.  Compare this to the
``worst case'' effective $A$-term contribution $\delta\lambda_{T,X_t}\sim 0.006$. 
To conclude, we find that even for relatively large $X^2_t<6m^2_{\tilde t}$, the
radiative corrections to the effective quartic coupling $\lambda(m_S)$ can be
negative, and can cancel the shift violating contributions in some scenarios.
At higher soft breaking scales, the relative importance of the $y_t^4$
corrections will dwindle, and the gaugino-higgsino thresholds will dominate the
corrections to $\lambda(m_S)$.  

\subsection{Models with $\lambda<0$}
\label{sec:negativelambda}
As we have discussed, the SM running of the quartic coupling leads us
into a regime where $\lambda<0$ far below the Planck scale if the Higgs/top 
quark masses turn out to be respectively at the lower/upper end of their 
current experimentally allowed ranges. It is hence interesting to 
analyze whether a supersymmetric/stringy UV completion can be realized 
in this classically unstable regime.

The naive answer is `no' since the MSSM $D$-term potential always implies
a non-negative tree-level quartic coupling at the soft scale (cf. the 
familiar tree-level relation (\ref{lambda})). In fact, we have so far 
mainly exploited the limiting case $\tan\beta=1$ and $\lambda(m_S)=0$, 
thereby pushing the SUSY breaking scale as far up as possible. Of course, 
$\lambda$ can receive loop corrections of either sign. Going beyond this, 
we will now appeal to classical corrections and argue that a {\it sizable} 
negative quartic coupling is a possibility that should be taken seriously. 

It is well-known that the upper bounds on the mass of the light Higgs in the
MSSM are loosened in extensions like the NMSSM, where the $F$-term potential 
of the singlet provides a contribution to the quartic coupling. Such effects 
can, of course, also be relevant in high-scale SUSY breaking. We will base
our discussion, once again, on the toy model (\ref{eq:ftermtoymodel}), also 
considered in \cite{Giudice:2011cg}. In other words, we introduce a singlet 
chiral superfield $S$, allowing for a renormalizable coupling to the doublets
and a supersymmetric mass term. 

While the addition of a singlet scalar to the Higgs sector may seem ad hoc
in field theory, it is quite natural in string models with soft breaking 
scales near the compactification scale. For example, if the chiral 
Higgs doublets $H_u, H_d$ originate from a hypermultiplet on intersecting 
branes as proposed e.g. in \cite{Ibanez:2001nd}, the bulk gauge theory 
degrees of freedom couple precisely in this manner\footnote{Such couplings to
SM singlets can also occur if $H_u$ and $H_d$ originate from separate matter
curves, in which case $S$ is charged and can become massive via instanton effects.}
\cite{higherdsuperspace, p-term}.  The degree of decoupling of the bulk
$F$-terms from the 4D effective theory is then a crucial issue (see also
section \ref{sec:constraintsonextendedsusy}). We have seen that a non-vanishing
quartic coupling will generally appear in the effective theory after
integrating out the singlet scalar if the latter has a nonvanishing soft mass
$m_S$. 

We now consider situations where $\tan\beta=1$ and, at the same 
time, $-M^2 < m_s^2 < 0$. As we will see, the main conclusion can be 
summarized in the equations
\be
{\cal W} = \kappa S H_u H_d + \frac{M}{2} S^2 + \dots\quad 
\stackrel{\tan\beta=1\,\,;\,\,m_s^2<0}{\Longrightarrow} 
\quad
V_{\Lambda<M}=\frac{\kappa^2 m_s^2}{M^2+m_s^2}|H_0|^4<0\,,\label{vlm}
\ee
where $H_0$ is the Standard Model Higgs doublet (which is massless at the 
high scale). The crucial point is that, in spite of the obvious tree-level 
instability, the running of $\lambda$ will save the theory in the expansion 
around $H_0=0$ (until, of course, at very low energies, the negative mass
squared of $H_0$ becomes relevant).

Let us now make this point in somewhat more detail, following the 
effective field theory from high to low energy scales. We will assume 
for simplicity that the Higgs mass entries $m_1^2=m_2^2=m_3^2$ are somewhat
(but not hierarchically) smaller than $M^2$ and $-m_s^2$, which are of the 
same order of magnitude. The full classical scalar potential reads
\begin{eqnarray}
V_{\Lambda>M}&=&|\kappa H_u H_d+MS|^2+|\kappa SH_u|^2+|\kappa SH_d|^2
+\left\{\kappa \overline{\mu}S(|H_u|^2+|H_d|^2)+\mbox{h.c.}\right\}\nonumber \\
&&+m_1^2|H_u+{H}_d^\dagger|^2+\frac{g_2^2+g_1^2}{8}(|H_u|^2-|H_d|^2)^2
+\frac{g^2}{2}|H_u\epsilon {H}_d^\dagger|^2+m_s^2 |S|^2
\end{eqnarray}
At energy scales high enough to neglect all masses, we have just the 
positive-definite quartic $D$ and $F$ term potentials and no flat directions 
are left. At smaller energies (smaller field vevs) one discovers an 
approximately flat direction. It corresponds to simultaneously switching on
$H_{u,d}$ and $S$. In fact, including even higher powers in the analysis, one 
sees that that the potential falls slightly below zero when moving from 
zero into this flat direction. After integrating out the superfield $S$, this 
behavior translates to a negative treelevel quartic coupling for the light 
Higgs doublet $H_0$ in the effective theory below $M$ (cf. (\ref{vlm})).

\begin{figure}
\begin{center}
\begin{picture}(220,150)
\put(-10,140){$V$}
\put(-5,85){$V_0$}
\put(18,45){EW Vacuum}
\put(95,5){true minimum}
\put(210,50){$H_0$}
\put(0,0){
\includegraphics[width=8cm]{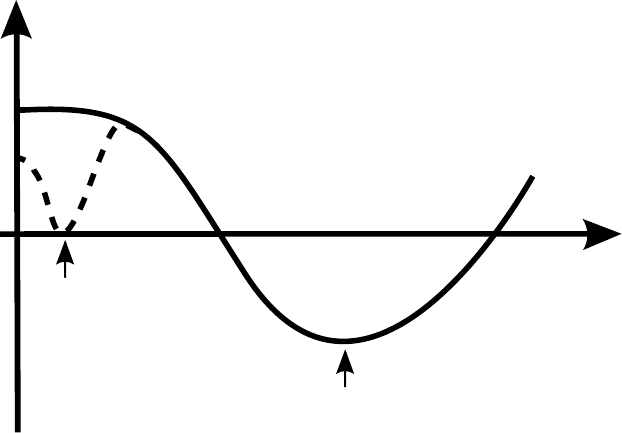}}
\end{picture}
\end{center}
\caption{
A sketch of the effective potential with negative quartic coupling before running (solid)
and after running to low energies (dashed). An uplift term to ensure vanishing cosmological constant in the electroweak (EW) vacuum has been added. 
\label{pot}}
\end{figure}
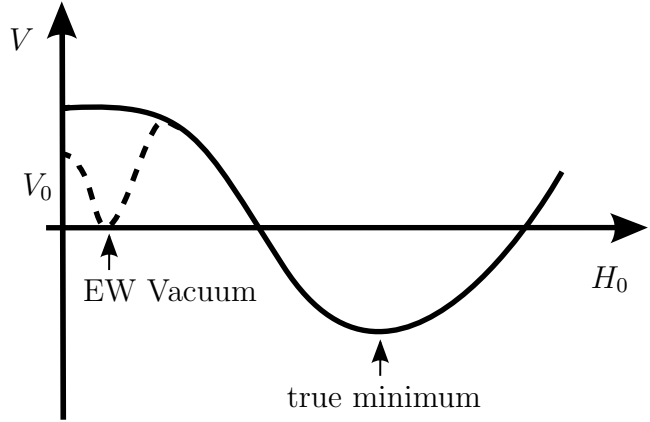

The potential at $H_0=0$ is unstable but we know (given the stability of our 
original softly broken SUSY model) that a stable minimum with all vevs 
non-zero exists. However, we propose to follow the RG flow of the theory 
near $H_0=0$. Our crucial point is that this is consistent and, once we reach 
the regime where $\lambda>0$, we just continue to run conventionally 
further down until standard electroweak symmetry breaking sets in. Of 
course, the previously discovered negative-energy minimum is still present 
and `our' minimum is only metastable. The situation is depicted in 
Fig~\ref{pot}. 

Before worrying about the consistency of our analysis, let us first see what 
we would gain. Clearly, we can significantly loosen the upper bound for the 
heavy MSSM masses in our model, at least if the top quark turns out to be 
relatively heavy. For example, for $m_t=173.5$ GeV, the 2-loop running gives 
us a minimum value of the SM quartic coupling in $\overline{MS}$ of 
$\lambda(10^{17.5}\mbox{ GeV})\approx -0.016$. Neglecting the running of the 
potential between $M$ and $m_S$, a negative correction of this size can be
obtained for $\kappa\approx 1$ and $M\approx 8 |m_s|$. Obviously, we can even shift the 
SUSY breaking scale all the way up to the Planck scale if we wish. 

But is this picture consistent? While the potential is unstable around $H_0=0$ 
due to the negative effective quartic coupling, this instability is not 
tachyonic. There is no (quadratic) mass instability or even tadpole. Thus, we 
encounter no technical problem in doing perturbation theory (with a
massless Higgs) around this extremum. The quartic instability introduces no
mass scale into our model and hence, given a sufficiently smooth vacuum state, 
we can `live' for a long time at $H_0=0$. The dynamics is governed by an 
effective field theory (with UV and IR cutoff), which can in principle be 
tested rather precisely (cf. Appendix \ref{rel} for more details).

This is, of course, in no way surprising from the perspective of cosmological 
inflation, where one is used to analysing potentials in perturbation theory 
around points with $V''<0$ (see, e.g., \cite{Guth:1982ec} for some early and 
more recent examples). If one also has $V'=0$, the lifetime of some 
Hubble-sized patch on the `top of the hill' is limited by the quantum 
fluctuations of the inflaton, which are controlled by the Hubble scale $H$. 
Obviously, there are also the familiar dS space IR divergences 
\cite{Vilenkin:1982wt} and, in general, late-time non-perturbative effects 
\cite{Starobinsky:1986fx}. However, all of this does not affect our 
application since, given the usual tuning of the cosmological constant, there 
is no high Hubble scale. Instead, as explained in the Appendix
\ref{apx:negativelambda}, we can alway use a low IR cutoff and this cutoff sets
the scale for both quantum diffusion and classical instability. 

All we really need for our purposes is to calculate how $\lambda$,
defined originally at a scale $\mu\sim M$, changes with $\mu$. The 
definition of this $\lambda$ is, very intuitively, via a 4-point correlation 
function at distances $\sim 1/\mu$ (hence without IR sensitivity) and with 
a UV cutoff $\Lambda_{UV}$ not too far above $\mu$. The answer to the 
question of the $\mu$-dependence of $\lambda$ is, of course, the conventional 
SM $\beta$-function. Thus, if we continue to calculate $\lambda$ at lower
and lower energy scales, its sign will eventually change and we return to 
the firm ground of perturbation theory around a stable extremum (again, 
obviously, before $\mu$ becomes comparable to the Higgs mass parameter). 

While we believe that the arguments given above (and in App. \ref{rel}) 
are sufficiently convincing at the leading-log level, a more careful field 
theoretic discussion is certainly worthwhile. In particular, it is 
interesting to investigate how far one can push higher-orders perturbation 
theory in this classically unstable regime. Another interesting question 
is how our universe has ended up in the radiatively generated, metastable 
state we just argued for. This may find a resolution along the lines suggested 
in \cite{Lebedev:2012sy} or in \cite{Riotto:1995am} (in a somewhat different 
but related contexts). 

We reiterate that, as we have just seen, a stringy UV completion might occur 
far above the scale where $\lambda$ turns negative. In the region with 
negative $\lambda$, we have no proper equilibrium field theory (in particular 
the Hamiltonian is unbounded below), but sufficient control in an effective 
theory with IR cutoff to answer `short-time-scale' questions. Most 
importantly, we control the running in this regime.

\section{Conclusions}

The discovery of a Higgs-like boson with a mass of
$\sim$ 126 GeV has provided a measurement of the last undetermined parameter of
the Standard Model: the quartic Higgs coupling $\lambda$. While it remains to be seen
whether the minimal Higgs mechanism truly is the correct theory of electroweak
symmetry breaking at and beyond the electroweak scale, all known properties of
the new Higgs-like particle such as decay branching fractions and production
rates are in agreement with SM predictions (see e.g.
\cite{Plehn:2012iz}). Absent new physics, the Higgs quartic
coupling in the SM generically runs to $\lambda=0$ and even negative values at
some high renormalization scale $\mu_\lambda \gtrsim 10^8 $ GeV for most of the
experimentally allowed range of top quark and Higgs mass values (see e.g. 
\cite{Lindner:1988ww,Cabrera:2011bi,Degrassi:2012ry,Holthausen:2011aa,Giudice:2011cg,EliasMiro:2011aa,Masina:2012tz}).  
For the purpose of this paper, we
have taken this (and the current lack of evidence for new physics at
colliders) as a hint that the scale $\mu_\lambda$ of vanishing quartic coupling might tell us
something about the nature of the UV completion of the SM. This general idea is
of course not new. There have for example been proposals of unified
theories \cite{Gogoladze:2007qm} and asymptotic safety scenarios
\cite{Shaposhnikov:2009pv} which have made predictions of the Higgs
mass based on the UV boundary condition $\lambda(Q)=0$ and
$\lambda(Q)=\beta_\lambda(Q)=0$ before the fact. 

We assume in this paper that superstring/M-theory is the correct description of
quantum gravity and fundamental interactions, and have identified models with
high-scale supersymmetry (for a review of alternatives see e.g.
\cite{Nath:2012nh}, for a recent discussion of the phenomenology of the
low intermediate regime see \cite{Arbey:2013jla}) which naturally exhibit approximately flat directions in
the Higgs potential, thus providing the UV boundary condition $\lambda(Q)=0$
for the RG running (as well as $m_h^2(Q)=0$ at tree level in the case of shift
symmetric models).  Most of our arguments hinge on the MSSM treelevel relation
$\lambda(\tan\beta=1)=0$ which results from the flat directions of the $D$-term potential.
In the decoupling limit in which we are working, the value $\tan\beta=1$ is
tied to a particular structure of the Higgs mass matrix where all entries are
equal with $B\mu = m_1^2=m_2^2$.  Such mass matrices occur automatically if the
Higgs sector features a shift symmetry \cite{Hebecker:2012qp} or an exchange symmetry
\cite{Ibanez:2012zg}. In the latter case, only $m_1^2=m_2^2$ is imposed by the
symmetry, and $B\mu\approx m^2_i$ is a by-product of the tuning of the electroweak
scale.

In our previous paper\cite{Hebecker:2012qp}, we have identified shift symmetric
Higgs sectors as a plausible explanation for $\tan\beta=1$ at a high scale.
While such shift symmetries have been known to appear in heterotic orbifold
models for Wilson line type bulk fields, it is unclear to what
degree they are preserved when going from orbifolds to smooth Calabi-Yaus.  In
this paper, we have focused on type II/F-Theory compactifications with D6 and
D7 branes. In analogy to the heterotic case, the obvious place where one would 
expect shift symmetries of the required type are again Wilson lines 
on D6 or D7 branes. It is important to note that the shift symmetry needs
to be realized in the correct K\"ahler variables of the 4D effective
supergravity theory in order to yield the desired soft masses. Not all
components of the Higgs field can exhibit a shift symmetry simultaneously (this
would forbid all terms in the K\"ahler potential). This, along with the
appearance of a shift-violating Chern-Simons term, prevents the desired
shift symmetry from being generically obeyed by D7 Wilson lines.  In the case on D6 branes, the
complex variables of the supersymmetric theory each combine one real Wilson
line degree of freedom with one brane scalar describing normal movement.  We have
argued that such bulk Higgs constructions and their type IIB duals (with a 
Higgs from D7 brane scalars) indeed exhibit shift symmetries of the
required type. 

A further possibility to implement shift/exchange symmetric models might be
given by Higgs sectors localized on matter curves of intersecting D7 branes. In
\cite{Ibanez:2012zg}, such scenarios were proposed as candidates for shift
symmetric models. The crucial question - what is the moduli dependence of the
K\"ahler metric pertaining to the $H_u H_d+c.c.$ terms - can not be answered in
a straightforward fashion by matching to the dimensionally reduced action. In
analogy to radion mediated SUSY breaking in 5D/6D models, we conjecture that
the relevant part of the K\"ahler metric might be recovered by considering
dimensionally reduced actions in the presence of warping. This may open up
the possibility to explicitely fine-tune the Higgs mass through fluxes.

We have then analyzed scenarios with an exchange symmetry realized
with D6 branes at angles\cite{Ibanez:2012zg}. A
crucial feature of these models is that the diagonal entries in the Higgs mass
matrix are generated solely via a supersymmetric mass term corresponding to a
small separation of the brane stacks supporting the Higgs hypermultiplets. Due
to the absence of soft Higgs masses at leading order, the scenario is
automatically exchange symmetric. The challenge is then to generate a $B\mu$
term of equal size without switching on comparable diagonal soft masses.  In
models \cite{Cremades:2002cs,Ibanez:2001nd} where D6 brane
configurations break SUSY completely (a situation which is dual to SUSY
breaking open string fluxes in IIB) such terms can be generated.  However, the
interpretation of a soft breaking parameter as an ``$F$-term like'' $B\mu$ term
rather than diagonal ``$D$-term like'' soft masses rests entirely on identifying
one particular 4D $\mathcal N=1$ supersymmetry as the surviving one. This
scheme only yields $\lambda=0$ in the low energy theory if the Higgs quartic
potential is precisely an MSSM-like $D$-term with respect to this surviving
generator. We have shown how this can in principle be achieved in the
simplified setting of D6 branes at angles in flat space. The key is that
certain types of brane angles (non-factorizable) can actually correspond
to $F$-terms.

Shift/exchange symmetry in the Higgs sector is necessarily broken at 1-loop
level due to the Yukawa couplings to matter, in particular to the top quark.
We have already given the resulting corrections to $\tan\beta=1$ in 
\cite{Hebecker:2012qp} and found that they are small enough to
maintain predictivity of an approximate shift/exchange symmetry. In this paper, we
have also considered the loop corrections to the quartic coupling itself. The
decoupling and threshold contributions which arise when the heavy MSSM
states are integrated out generically are of similar size as the shift violating corrections.
Both types of corrections merely yield shifts of the physical Higgs mass of
$|\Delta m_h|\lsim 2$ GeV, often less.  This is in contrast to TeV scale
SUSY and is mainly due to the relative smallness of the top Yukawa coupling at
high renormalization scales\cite{Hall:2009nd}. Furthermore, 
in scenarios with a small splitting
between the soft parameters, some of the contributions tend to cancel. 
In summary, we find that in
all but the extreme cases, the overall loop corrections to the quartic
coupling are small, and the preferred region for the soft breaking scale
remains close to the naive estimate.

Finally, we have considered whether the incomplete decoupling of $F$-term
potentials (which we have tried to avoid until now) might in some situations
provide us with a UV completion with $\lambda<0$. We have shown, based on a
simple NMSSM-like model, how such an unusual situation can arise. We found that
the soft scale at which our UV completion comes in can be raised all the way up
near the Planck scale even if the top quark mass turns out to be towards the
heavy end of the currently favored range.  
While one naively expects that some new physics must come into play at
$\mu_\lambda$ to avoid the instability of the scalar potential, this is not
strictly speaking necessary: There may simply be an energy gap between
$\mu_\lambda$ and $m_{UV}$ (e.g. the string scale) where no stable 4d effective
theory exists. Also, arguing for new physics at $\mu_\lambda$ from a
cosmological point of view is not fully convincing: In the early history of the
universe, even if it was very hot, the finite-temperature effective potential
will presumably be stable independently of the running of $\lambda$. In the
late history, all that matters is a sufficiently long lifetime of our (possibly
metastable) minimum\cite{Degrassi:2012ry}. In order to match the SM running of
the quartic coupling to such an unstable UV completion, we have addressed a
different but related question, namely whether perturbation theory in the
$\lambda<0$ regime is stable enough such that $\lambda$ is still a meaningful
UV boundary conditions of the SM RGEs. We have given some simple arguments
which suggest that the lifetime of ``vacuum'' states is sufficiently long to allow
such a matching at least at the level of precision required for our purposes.

In this paper, we have not discussed dark matter nor flavor physics in any
detail. As we have already noted in \cite{Hebecker:2012qp}, and as was worked
out in \cite{Ibanez:2012zg}, the typical scales which appear in
these high-scale SUSY scenarios are in an interesting range for Axion CDM as well
as near typical Seesaw scales.
Indeed, while it does not seem impossible to extend our models to variations of
split SUSY \cite{ArkaniHamed:2004fb} with an electroweak WIMP candidate at the
TeV scale (however, see \cite{Arvanitaki:2012ps} for some Caveats), axions
are ubiquitous in string compactifications and certainly provide the most
compelling dark matter candiates in this class of models. Further research in
this direction would certainly be interesting. Other recent proposals relating
axions to high-scale physics can for example be found in
\cite{Chatzistavrakidis:2012bb,Redi:2012ad}. An interesting alternative to the
MSSM or NMSSM like Higgs sectors in high-scale SUSY are single-Higgs SUSY
models (see \cite{Davies:2011mp,Unwin:2012fj}). For models employing shift symmetries
of the type discussed in this paper in the context of LARGE volume scenarios, see
\cite{Higaki:2012ar}.

\section*{Acknowledgments}
We would like to thank M. Drees, M. Goodsell, T. Grimm, H. Jockers, A. M\"uck, A. Pomarol and J. Unwin for useful discussions.
This work was partially supported by the Transregio TR33 "The Dark Universe".
\section*{Appendix}
\appendix

\section{The transition from bulk matter to intersection matter}
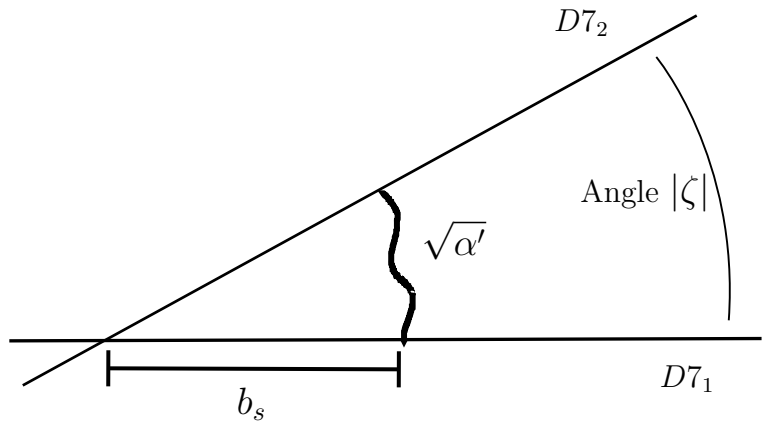
\begin{figure}[h]
\begin{center}
\begin{picture}(300,150)
\put(250,0){$D7_1$}
\put(210,135){$D7_2$}
\put(220,70){Angle \large $|\zeta|$}
\put(90,-10){\large $b_s$}
\put(160,50){\large $\sqrt{\alpha'}$}
\put(0,0){
\includegraphics[width=10cm]{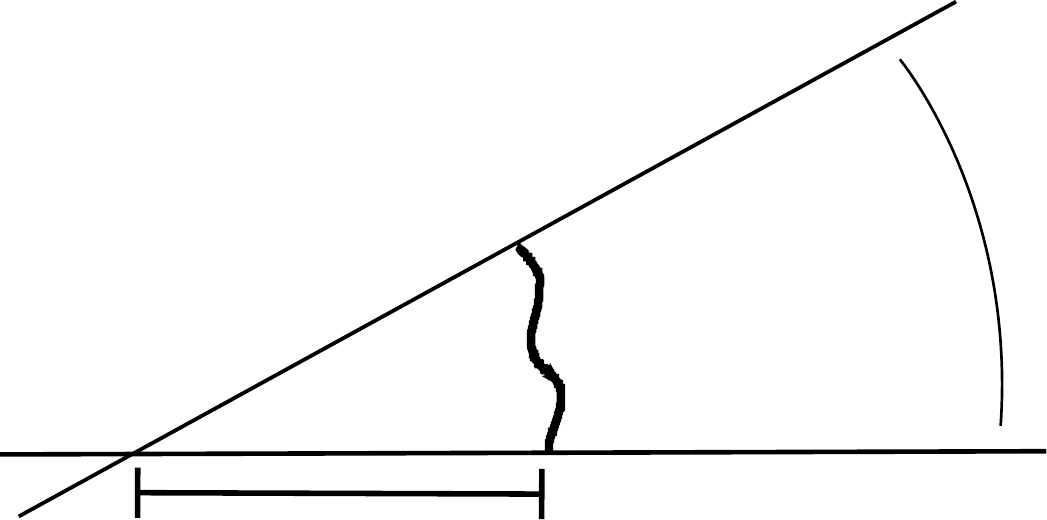}}
\end{picture}
\end{center}
\caption{An illustration of the scales involved in the localization of string states in the bifundamental representation.\label{localizationillu}}
\end{figure}
In section \ref{sec:typeIImodels} we discuss the possibilities for shift
symmetry in IIB models in which the Higgs sector is realized as D7 brane
matter. We now want to see how matter on intersection curves arises as a
deformation of this scenario in order to study the moduli dependence of the
matter K\"ahler potential.

There is in principle a continuous transition from localized matter to matter
on a brane stack if the compactification allows a brane deformation
modulus $\zeta$ corresponding to a nontrivial profile which locally looks like
$\phi \sim |\zeta| z_1$ of one of the adjoint brane scalars $\phi$ parameterizing the
relative transverse brane movement (here, $z_1$ is one of the internal brane-parallel
 directions $z_1,z_2$, and the normalization of $\zeta$ is such that
$\zeta\sim\mathcal O(1)$ corresponds to generic angles).
However, the known moduli dependences of
$K$ on the dilaton $S$ and K\"ahler moduli $T_i$ differ between the two cases\cite{Jockers:2004yj,Aparicio:2008wh}.
It is thus our aim to understand at which point 
this transition between a ``bulk-like'' and  a ``matter-curve-like'' situation occurs.

For simplicity we assume two spacetime filling D7 branes with parallel directions $z_1$ and $z_2$ in the extra dimensions. 
Our starting point is a situation where they are on top of each other with the known bulk-matter K\"ahler metric\cite{Jockers:2004yj,Aparicio:2008wh,Conlon:2006tj,Font:2009gq,Kawano:2011aa} 
\begin{equation}K_{Bulk}\sim s^{-1}\end{equation} and no dependence on $t$, where $s$ and $t$ are the real parts of the dilaton and relevant K\"ahler modulus respectively. We now switch on an adjoint vev 
\begin{equation}
\langle\phi\rangle \sim \gamma z_1
\end{equation}
where the constant $\gamma$ parameterizes the angle. The brane-matter coupled to this vev profile now also obtains a nontrivial profile
\begin{equation}
\psi(z_1,z_2)\sim e^{-\gamma|z_1|^2 } \, f(z_2)\,.
\end{equation}
Integrating over $z_1$ and $z_2$ in the kinetic term, we obtain
\begin{equation}
\int \! d^2 z_1 \int \! d^2 z_2 |\psi|^2 = \int\! d^2 z_2 |f|^2 \, \int \! d^2 z_1 e^{-\gamma |z_1|^2}\sim \frac{1}{\gamma}
\end{equation}
Consequently, the kinetic part of the matter K\"ahler potential derived from the intersection curve scales like
\begin{equation}
K_{int} \sim s^{-1} \gamma^{-1}\,.
\end{equation}
We thus have to determine the dependence of $\gamma$ on the moduli.
Let us first note that the typical extension of the ``classical'' particle wave functions in terms of complex geometry coordinates is
\begin{equation}
|z^{crit}| \sim \gamma^{-1/2}\,.
\end{equation}
We now match this field theory quantity to the string picture illustrated in Figure \ref{localizationillu}.
It is clear that the extension of the particle wave functions is given by the 
region of size $b_s$ where the brane separation is small compared to the string length. In the string frame,
the relation between $z^{crit}$ and $b_s$ is, up to proportionality factors, given by
\begin{equation}b_s \sim |z_{crit}| R_s\end{equation}
Here, $R_s$ is the typical radius of the compactification in the string frame which scales like the K\"ahler modulus.
In the string frame, the string length is constant, and we can thus read off (see Fig. \ref{localizationillu}) that
\begin{equation}
|\zeta| b_s \sim \sqrt{\alpha'}
\end{equation}
and consequently
\begin{equation}
\gamma^{-1/2}R_s \sim b_s \sim \frac{\sqrt{\alpha'}}{|\zeta|}\sim \frac{1}{|\zeta|}\,.
\end{equation}
The relation between string frame and Einstein frame metric is given by
\begin{equation}
g_{ij}^s=g^E_{ij}g_s^{1/2}
\end{equation}
and thus likewise for the K\"ahler modulus
\begin{equation}
R_s^2 \sim R_E^2 g_s^{1/2}\sim t^{1/2}\, g_s^{1/2}=t^{1/2}s^{-1/2}\,.
\end{equation}
Putting everything together, we find that
\begin{equation}
K_{int}\sim \frac{1}{s}\frac{1}{\gamma} =\frac{1}{s}\frac{1}{R_s^2 |\zeta|^2} = \frac{1}{\sqrt{st}}\,\frac{1}{|\zeta|^2}
\end{equation}
This moduli dependence is consistent with what was found in
\cite{Aparicio:2008wh}. We also obtain the dependence on the brane deformation
modulus $\zeta$.

When does the transition between bulk and intersection curve K\"ahler potential occur? We
have assumed that the extent of the classical wave function is limited by the
brane separation. In this situation, the wave function does not ``feel'' the
overall volume of the cycle on which the brane is wrapped. This changes as soon
as the extent of the classical wave function becomes comparable to the
compactification radius, at $\gamma=1$ or $|\zeta|^2 \sim \sqrt{s/t}$. 
This transition is captured in (\ref{qua}).

\section{Reliability of effective field theory in the classically 
\label{apx:negativelambda}
unstable regime of negative $\lambda$}\label{rel}

In section \ref{sec:negativelambda} we discuss UV completions with negative
quartic coupling and argue that it is possible to match them perturbatively to
the SM at the soft breaking scale. Let us try to be a bit more precise: We
consider a model with a massless scalar $H_0$ with negative quartic potential.
This is not the full theory, but a good approximation as long as $|H_0|\ll M$.
Furthermore, we pretend for the sake of the following qualitative analysis that
$H_0$ is a single real scalar rather than a complex doublet. 

To get rid of IR problems, we compactify the theory on a $T^3$ with volume 
$R^3$. The instability is associated only with the zero mode $\phi$ of $H_0$. 
Neglecting for the moment the quartic coupling, the dynamics of this mode 
corresponds simply to that of a free quantum mechanical particle with 
position $x=\phi R^2$ and mass $m=1/R$,
\be
S=\int dt\, R^3 \frac{1}{2}\dot{\phi}^2=\int dt \frac{1}{2R}\dot{x}^2\,.
\ee
Let us take our initial state to be a Gaussian Schr\"odinger wave function 
with width $\delta x_0$ and momentum uncertainty $\sim 1/\delta x_0$. According 
to the Schr\"odinger equation, it will spread with time in the familiar way: 
$\delta x\sim \sqrt{\delta x_0^2+(t-t_0)^2/4m^2\delta x_0^2}$. Returning to the 
field theory model, this translates into an unavoidable quantum uncertainty 
of the field position $\delta \phi$ and field velocity $\delta\dot{\phi}\sim 
1/(R^3\delta\phi_0)$. The former then grows as 
\be
\delta\phi \sim \sqrt{\delta\phi_0^2 + (t-t_0)^2/4R^6\delta\phi_0^2}\,,
\label{qdiff}
\ee
while the latter remains fixed. 

We want to match this to the classical analysis of vacuum decay in the 
inverted quartic potential:
\be
\rho=\frac{\dot{\phi}^2}{2}-|\lambda|\phi^4 \quad \Rightarrow \quad 
t=\int\frac{d\phi}{\sqrt{2(\rho+|\lambda|\phi^4)}}\,.\label{rhodef}
\ee
Here $\rho = E/R^3$ is the (conserved) energy density associated with the 
zero mode. One immediately observes two regimes: An early regime where
\be
\phi = \phi_1 + (t-t_0) \sqrt{2\rho}\,,\label{cevol}
\ee 
and a late regime, 
\be 
\phi = \frac{1}{\sqrt{2|\lambda|}(-t)}\qquad\mbox{with}\qquad t<0\,,
\ee
diverging in finite time as expected. The matching is at the time or, 
equivalently, $\phi$-value satisfying $|\lambda|\phi^4\sim \rho$, with
$\rho$ a conserved quantity encoding the initial conditions. 

We could now optimize our choice of quantum state (i.e. of $\delta\phi_0$) 
to live for as long as possible `on the top of the hill'. Instead, we start 
with a simple-minded guess: Let $\delta\phi_0 \sim 1/R$, such that $R$ 
remains the only dimensionful quantity in the problem, and attempt to match 
this immediately to the classical description. According to our Gaussian 
wave function\footnote{
Actually, 
we are of course dealing with a wave functional, but we can ignore the 
non-zero modes for our purposes.
}
the typical classical field configuration has $\phi_1\sim \delta \phi_0 
\sim \pm 1/R$, velocity $\dot{\phi}\sim 1/R^2$ and energy density $\rho \sim 
1/R^4$ (cf. (\ref{rhodef}), assuming also $|\lambda|\ll 1$). The field 
evolves classically (cf. (\ref{cevol}) with $t_0=0$) as 
\be
\phi \sim \pm \frac{1}{R}+\frac{t}{R^2}\,,
\ee
reaching the dangerous late regime of fast decay at $\phi\sim 
1/(|\lambda|^{1/4}R)$. The corresponding critical time is $t_c \sim R/
|\lambda|^{1/4}$. Our early matching from quantum to classical is a 
posteriori justified by the observation that the same conclusion would have 
been reached in quantum evolution according to (\ref{qdiff}). 

Thus, we can safely think in terms of a theory with IR cutoff $1/R$, asking any
dynamical question which does not require time scales larger than $t_c$ (with
$t_c\gg R$ for $|\lambda| \ll 1$). The cutoff can be chosen as low as we wish.
It will generically induce relevant operators such as an effective Higgs mass,
but since these are suppressed relative to $1/R^2$ by a loop factor, they should
not interfere with physics at time scales relevant for our argument.  Thus, our
scenario with negative high-scale $\lambda$ works as long as $\lambda$ runs to
a positive value somewhere in the IR.


\begin{thebibliography}{99}
\bibitem{expatlas} 
  G.~Aad {\it et al.}  [ATLAS Collaboration],
  ``Observation of a new particle in the search for the Standard Model Higgs boson with the ATLAS detector at the LHC,''
  Phys.\ Lett.\ B {\bf 716}, 1 (2012)
  [arXiv:1207.7214 [hep-ex]].\\
For recent updates, see\\
The ATLAS collaboration, ATLAS-CONF-2013-014\\
The ATLAS collaboration, ATLAS-CONF-2013-034.
\bibitem{expcms} 
  S.~Chatrchyan {\it et al.}  [CMS Collaboration],
  ``Observation of a new boson at a mass of 125 GeV with the CMS experiment at the LHC,''
  Phys.\ Lett.\ B {\bf 716}, 30 (2012)
  [arXiv:1207.7235 [hep-ex]].\\
For recent updates (in particular of the $\gamma\gamma$ channel), see\\
The CMS collaboration, CMS PAS-HIG-12-045\\
The CMS collaboration, CMS PAS-HIG-13-001.

\bibitem{Plehn:2012iz} 
  T.~Plehn and M.~Rauch,
  ``Higgs Couplings after the Discovery,''
  Europhys.\ Lett.\  {\bf 100}, 11002 (2012)
  [arXiv:1207.6108 [hep-ph]].\\
  A.~Djouadi and G.~ég.~Moreau,
  ``The couplings of the Higgs boson and its CP properties from fits of the signal strengths and their ratios at the 7+8 TeV LHC,''
  arXiv:1303.6591 [hep-ph].

\bibitem{Baak:2011ze}
  M.~Baak, M.~Goebel, J.~Haller, A.~Hoecker, D.~Ludwig, K.~Moenig, M.~Schott
  and  J.~Stelzer,
  ``Updated Status of the Global Electroweak Fit and Constraints on New 
  Physics,''
  Eur.\ Phys.\ J.\ C {\bf 72} (2012) 2003
  [arXiv:1107.0975 [hep-ph]].
\bibitem{Isidori:2010kg}
  G.~Isidori, Y.~Nir and G.~Perez,
  ``Flavor Physics Constraints for Physics Beyond the Standard Model,''
  Ann.\ Rev.\ Nucl.\ Part.\ Sci.\  {\bf 60} (2010) 355
  [arXiv:1002.0900 [hep-ph]].
\bibitem{Denef:2004ze}
  F.~Denef and M.~R.~Douglas,
  ``Distributions of flux vacua,'' JHEP {\bf 0405} (2004) 072
  [hep-th/0404116]  and ``Distributions of nonsupersymmetric flux vacua,'' 
  JHEP {\bf 0503} (2005) 061 [hep-th/0411183].
\bibitem{ArkaniHamed:2004fb}
  N.~Arkani-Hamed and S.~Dimopoulos,
  ``Supersymmetric unification without low energy supersymmetry and 
  signatures for fine-tuning at the LHC,'' JHEP {\bf 0506} (2005) 073
  [hep-th/0405159];\\
  G.~F.~Giudice and A.~Romanino,
  ``Split supersymmetry,''
  Nucl.\ Phys.\ B {\bf 699} (2004) 65
   [Erratum-ibid.\ B {\bf 706} (2005) 65]
  [hep-ph/0406088];\\
  L.~J.~Hall and Y.~Nomura,
  ``Spread Supersymmetry,''
  JHEP {\bf 1201} (2012) 082
  [arXiv:1111.4519 [hep-ph]];\\
  N.~Arkani-Hamed, A.~Gupta, D.~E.~Kaplan, N.~Weiner and T.~Zorawski,
  ``Simply Unnatural Supersymmetry,''
  arXiv:1212.6971 [hep-ph].
\bibitem{Lindner:1988ww}
  M.~Lindner, M.~Sher and H.~W.~Zaglauer,
  ``Probing Vacuum Stability Bounds at the Fermilab Collider,''
  Phys.\ Lett.\ B {\bf 228} (1989) 139;\\
  M.~Sher,
  ``Electroweak Higgs Potentials and Vacuum Stability,''
  Phys.\ Rept.\  {\bf 179} (1989) 273;\\
  J.~A.~Casas, J.~R.~Espinosa and M.~Quiros,
  ``Improved Higgs mass stability bound in the Standard Model and 
  implications f  or supersymmetry,'' Phys.\ Lett.\ B {\bf 342} (1995) 171
  [hep-ph/9409458] and 
  ``Standard Model stability bounds for new physics within LHC reach,''
  Phys.\ Lett.\ B {\bf 382} (1996) 374 [hep-ph/9603227];\\
  C.~D.~Froggatt and H.~B.~Nielsen,
  ``Standard Model criticality prediction: Top mass 173 +- 5-GeV and Higgs 
  mass 135 +- 9-GeV,'' Phys.\ Lett.\ B {\bf 368} (1996) 96 [hep-ph/9511371];\\
  C.~D.~Froggatt, H.~B.~Nielsen and Y.~Takanishi,
  ``Standard Model Higgs boson mass from borderline metastability of the 
  vacuum,'' Phys.\ Rev.\ D {\bf 64} (2001) 113014 [hep-ph/0104161]. 

\bibitem{Shaposhnikov:2009pv}
  M.~Shaposhnikov and C.~Wetterich,
  ``Asymptotic safety of gravity and the Higgs boson mass,''
  Phys.\ Lett.\ B {\bf 683} (2010) 196 [arXiv:0912.0208 [hep-th]].

\bibitem{Cabrera:2011bi} 
  M.~E.~Cabrera, J.~A.~Casas and A.~Delgado,
  ``Upper Bounds on Superpartner Masses from Upper Bounds on the Higgs 
  Boson Mass,''
  Phys.\ Rev.\ Lett.\  {\bf 108}, 021802 (2012)
  [arXiv:1108.3867 [hep-ph]].

\bibitem{Giudice:2011cg}
  G.~F.~Giudice and A.~Strumia,
  ``Probing High-Scale and Split Supersymmetry with Higgs Mass Measurements,''
  Nucl.\ Phys.\ B {\bf 858} (2012) 63 [arXiv:1108.6077v2 [hep-ph]].

\bibitem{Holthausen:2011aa}
  M.~Holthausen, K.~S.~Lim and M.~Lindner,
  ``Planck scale Boundary Conditions and the Higgs Mass,''
  arXiv:1112.2415 [hep-ph];\\
  C.~Wetterich,
  ``Where to look for solving the gauge hierarchy problem?,''\\
  arXiv:1112.2910 [hep-ph].

\bibitem{EliasMiro:2011aa}
  J.~Elias-Miro, J.~R.~Espinosa, G.~F.~Giudice, G.~Isidori, A.~Riotto and 
  A.~Strumia,
  ``Higgs mass implications on the stability of the electroweak vacuum,''
  arXiv:1112.3022 [hep-ph].

\bibitem{Degrassi:2012ry}
  G.~Degrassi, S.~Di Vita, J.~Elias-Miro, J.~R.~Espinosa, G.~F.~Giudice, 
  G.~Isidori and A.~Strumia,
  ``Higgs mass and vacuum stability in the Standard Model at NNLO,''
  JHEP {\bf 1208} (2012) 098 [arXiv:1205.6497 [hep-ph]].

\bibitem{Masina:2012tz}
  I.~Masina,
  ``The Higgs boson and Top quark masses as tests of Electroweak Vacuum Stability,''
  Physical Review D 87, {\bf 053001} (2013)
  [arXiv:1209.0393 [hep-ph]].

\bibitem{threeloop}
  K.~G.~Chetyrkin and M.~F.~Zoller,
  ``$\beta$-function for the Higgs self-interaction in the Standard Model at three-loop level,''
  arXiv:1303.2890 [hep-ph];\\
  M.~F.~Zoller,
  ``Vacuum stability in the SM and the three-loop $\beta$-function for the Higgs self-interaction,''
  arXiv:1209.5609 [hep-ph];\\
  L.~N.~Mihaila, J.~Salomon and M.~Steinhauser,
  ``Renormalization constants and beta functions for the gauge couplings of the Standard Model to three-loop order,''
  Phys.\ Rev.\ D {\bf 86}, 096008 (2012)
  [arXiv:1208.3357 [hep-ph]];\\
  K.~G.~Chetyrkin and M.~F.~Zoller,
  ``Three-loop $\beta$-functions for top-Yukawa and the Higgs self-interaction in the Standard Model,''
  JHEP {\bf 1206}, 033 (2012)
  [arXiv:1205.2892 [hep-ph]];\\
  F.~Bezrukov, M.~Y.~.Kalmykov, B.~A.~Kniehl and M.~Shaposhnikov,
  ``Higgs Boson Mass and New Physics,''
  JHEP {\bf 1210} (2012) 140
  [arXiv:1205.2893 [hep-ph]];\\
  L.~N.~Mihaila, J.~Salomon and M.~Steinhauser,
  ``Gauge Coupling Beta Functions in the Standard Model to Three Loops,''
  Phys.\ Rev.\ Lett.\  {\bf 108}, 151602 (2012)
  [arXiv:1201.5868 [hep-ph]].

\bibitem{Martin:1997ns}
 For a review see e.g. S.~P.~Martin,
  ``A Supersymmetry primer,''
  In *Kane, G.L. (ed.): Perspectives on supersymmetry II* 1-153
  [hep-ph/9709356].
\bibitem{Hall:2009nd}
  L.~J.~Hall and Y.~Nomura,
  ``A Finely-Predicted Higgs Boson Mass from A Finely-Tuned Weak Scale,''
  JHEP {\bf 1003} (2010) 076
  [arXiv:0910.2235 [hep-ph]];
\bibitem{Hebecker:2012qp}
  A.~Hebecker, A.~K.~Knochel and T.~Weigand,
  ``A Shift Symmetry in the Higgs Sector: Experimental Hints and Stringy 
  Realizations,'' JHEP {\bf 1206} (2012) 093
  [arXiv:1204.2551 [hep-th]].
\bibitem{LopesCardoso:1994is}
  G.~Lopes Cardoso, D.~L\"ust and T.~Mohaupt,
  ``Moduli spaces and target space duality symmetries in (0,2) Z(N) orbifold
  theories with continuous Wilson lines,''
  Nucl.\ Phys.\  B {\bf 432} (1994) 68
  [arXiv:hep-th/9405002];\\
  I.~Antoniadis, E.~Gava, K.~S.~Narain and T.~R.~Taylor,
  ``Effective mu term in superstring theory,''
  Nucl.\ Phys.\  B {\bf 432} (1994) 187
  [arXiv:hep-th/9405024];\\
  A.~Brignole, L.~E.~Ib\'a\~nez, C.~Mu\~noz and C.~Scheich,
  ``Some Issues In Soft Susy Breaking Terms From Dilaton / Moduli Sectors,''
  Z.\ Phys.\  C {\bf 74} (1997) 157
  [arXiv:hep-ph/9508258];\\
  A.~Brignole, L.~E.~Ib\'a\~nez and C.~Mu\~noz,
  ``Orbifold-induced mu term and electroweak symmetry breaking,''
  Phys.\ Lett.\  B {\bf 387} (1996) 769
  [arXiv:hep-ph/9607405];
\bibitem{Brignole:1997dp}
  A.~Brignole, L.~E.~Ib\'a\~nez and C.~Mu\~noz,
  ``Soft supersymmetry-breaking terms from supergravity and superstring
  models,''
  arXiv:hep-ph/9707209.
  L.~J.~Dixon, V.~Kaplunovsky and J.~Louis,
  ``On Effective Field Theories Describing (2,2) Vacua of the Heterotic
  String,''
  Nucl.\ Phys.\  B {\bf 329}, 27 (1990).

\bibitem{Inoue:1985cw}
  K.~Inoue, A.~Kakuto and H.~Takano,
  ``Higgs as (Pseudo)Goldstone Particles,''
  Prog.\ Theor.\ Phys.\  {\bf 75} (1986) 664;\\
  A.~A.~Anselm and A.~A.~Johansen,
  ``SUSY GUT with Automatic Doublet - Triplet Hierarchy,''
  Phys.\ Lett.\ B {\bf 200} (1988) 331;\\
  Z.~G.~Berezhiani and G.~R.~Dvali,
  ``Possible solution of the hierarchy problem in supersymmetrical grand unification theories,''
  Bull.\ Lebedev Phys.\ Inst.\  {\bf 5} (1989) 55
   [Kratk.\ Soobshch.\ Fiz.\  {\bf 5} (1989) 42];\\
  R.~Barbieri, G.~R.~Dvali and A.~Strumia,
  ``Grand unified supersymmetric Higgs bosons as pseudoGoldstone particles,''
  Nucl.\ Phys.\ B {\bf 391} (1993) 487;\\
  H.~-C.~Cheng,
  ``Doublet triplet splitting and fermion masses with extra dimensions,''
  Phys.\ Rev.\ D {\bf 60} (1999) 075015
  [hep-ph/9904252];\\
  G.~Burdman and Y.~Nomura,
  ``Unification of Higgs and gauge fields in five-dimensions,''
  Nucl.\ Phys.\ B {\bf 656} (2003) 3
  [hep-ph/0210257].

\bibitem{Gogoladze:2007qm}
   I.~Gogoladze, N.~Okada and Q.~Shafi,
   ``Higgs boson mass from gauge-Higgs unification,''
   Phys.\ Lett.\ B {\bf 655} (2007) 257 [arXiv:0705.3035 [hep-ph]] and 
   ``Window For Higgs Boson Mass From Gauge-Higgs Unification,''
   Phys.\ Lett.\ B {\bf 659} (2008) 316 [arXiv:0708.2503 [hep-ph]].

\bibitem{Redi:2012ad}
  M.~Redi and A.~Strumia,
  ``Axion-Higgs Unification,''
  JHEP {\bf 1211} (2012) 103
  [arXiv:1208.6013 [hep-ph]].

\bibitem{BenDayan:2010yz}
  I.~Ben-Dayan and M.~B.~Einhorn,
  ``Supergravity Higgs Inflation and Shift Symmetry in Electroweak Theory,''
  JCAP {\bf 1012} (2010) 002
  [arXiv:1009.2276 [hep-ph]].

\bibitem{Unwin:2012fj}
  J.~Unwin,
  ``R-symmetric High Scale Supersymmetry,''
  Phys.\ Rev.\ D {\bf 86} (2012) 095002
  [arXiv:1210.4936 [hep-ph]].
\bibitem{Ibanez:2012zg}
  L.~E.~Ib\'a\~nez, F.~Marchesano, D.~Regalado and I.~Valenzuela,
  ``The Intermediate Scale MSSM, the Higgs Mass and F-theory Unification,''
  JHEP {\bf 1207} (2012) 195
  [arXiv:1206.2655 [hep-ph]].
\bibitem{Ibanez:2013gf}
  L.~E.~Ib\'a\~nez and I.~Valenzuela,
  ``The Higgs Mass as a Signature of Heavy SUSY,''
  arXiv:1301.5167 [hep-ph].
\bibitem{EliasMiro:2012ay} 
  J.~Elias-Miro, J.~R.~Espinosa, G.~F.~Giudice, H.~M.~Lee and A.~Strumia,
  ``Stabilization of the Electroweak Vacuum by a Scalar Threshold Effect,''
  JHEP {\bf 1206}, 031 (2012)
  [arXiv:1203.0237 [hep-ph]].\\
  L.~A.~Anchordoqui, I.~Antoniadis, H.~Goldberg, X.~Huang, D.~L\"ust, T.~R.~Taylor and B.~Vlcek,
  ``Vacuum Stability of Standard Model$^{++}$,''
  JHEP {\bf 1302}, 074 (2013)
  [arXiv:1208.2821 [hep-ph]].

\bibitem{Cicoli:2013rwa} 
  M.~Cicoli, S.~de Alwis and A.~Westphal,
  ``Heterotic Moduli Stabilization,''\\
  arXiv:1304.1809 [hep-th].

\bibitem{Buchmuller:2005jr}
  W.~Buchm{\" u}ller, K.~Hamaguchi, O.~Lebedev and M.~Ratz,
  ``Supersymmetric Standard Model from the heterotic string,''
  Phys.\ Rev.\ Lett.\  {\bf 96}, 121602 (2006)
  [arXiv:hep-ph/0511035].\\
  O.~Lebedev, H.~P.~Nilles, S.~Raby, S.~Ramos-Sanchez, M.~Ratz,
  P.~K.~S.~Vaudrevange and A.~Wingerter,
  ``A mini-landscape of exact MSSM spectra in heterotic orbifolds,''
  Phys.\ Lett.\  B {\bf 645} (2007) 88
  [arXiv:hep-th/0611095].
\bibitem{Choi:2003kq} 
  K.~w.~Choi, N.~y.~Haba, K.~S.~Jeong, K.~i.~Okumura, Y.~Shimizu and 
  M.~Yamaguchi,
  ``Electroweak symmetry breaking in supersymmetric gauge-Higgs  unification
  models,''
  JHEP {\bf 0402} (2004) 037
  [arXiv:hep-ph/0312178];\\
  A.~Hebecker, J.~March-Russell and R.~Ziegler,
  ``Inducing the mu and the B mu Term by the Radion and the 5d Chern-Simons 
  Term,'' JHEP {\bf 0908} (2009) 064 [arXiv:0801.4101 [hep-ph]];\\
  F.~Br{\"u}mmer, S.~Fichet, A.~Hebecker and S.~Kraml,
  ``Phenomenology of Supersymmetric Gauge-Higgs Unification,''
  JHEP {\bf 0908} (2009) 011 [arXiv:0906.2957 [hep-ph]].
\bibitem{Brummer:2010fr}
  F.~Br{\"u}mmer, R.~Kappl, M.~Ratz and K.~Schmidt-Hoberg,
  ``Approximate R-symmetries and the mu term,''
  JHEP {\bf 1004} (2010) 006
  [arXiv:1003.0084 [hep-th]];\\
  F.~Br{\"u}mmer, S.~Fichet, S.~Kraml and R.~K.~Singh,
  ``On SUSY GUTs with a degenerate Higgs mass matrix,''
  JHEP {\bf 1008}, 096 (2010)
  [arXiv:1007.0321 [hep-ph]].
\bibitem{Pena:2012ki} 
  D.~K.~M.~Pena, H.~P.~Nilles and P.~-K.~Oehlmann,
  ``A Zip-code for Quarks, Leptons and Higgs Bosons,''
  JHEP {\bf 1212}, 024 (2012)
  [arXiv:1209.6041 [hep-th]].


\bibitem{Ibanez:1987xa}
  L.~E.~Ib\'a\~nez, H.~P.~Nilles and F.~Quevedo,
  ``Reducing The Rank Of The Gauge Group In Orbifold Compactifications Of The
  Heterotic String,''
  Phys.\ Lett.\  B {\bf 192}, 332 (1987).
\bibitem{Aldazabal:1998mr}
  G.~Aldazabal, A.~Font, L.~E.~Ib\'a\~nez and G.~Violero,
  ``D = 4, N=1, type IIB orientifolds,''
  Nucl.\ Phys.\ B {\bf 536} (1998) 29
  [hep-th/9804026].\\
  R.~Blumenhagen, V.~Braun, T.~W.~Grimm and T.~Weigand,
  ``GUTs in Type IIB Orientifold Compactifications,''
  Nucl.\ Phys.\ B {\bf 815} (2009) 1
  [arXiv:0811.2936 [hep-th]].
\bibitem{Jockers:2004yj}
  H.~Jockers and J.~Louis,
  ``The Effective action of D7-branes in N = 1 Calabi-Yau orientifolds,''
  Nucl.\ Phys.\ B {\bf 705} (2005) 167 [hep-th/0409098].
\bibitem{Hebecker:2012aw}
  A.~Hebecker, S.~C.~Kraus, M.~Kuntzler, D.~L\"ust and T.~Weigand,
  ``Fluxbranes: Moduli Stabilisation and Inflation,''
  JHEP {\bf 1301} (2013) 095 [arXiv:1207.2766 [hep-th]].
\bibitem{Arends}
  M. Arends, A. Hebecker, K. Heimpel, S. C. Kraus, D. L\"ust, C. Mayrhofer, 
  C. Schick, and T. Weigand, ``Flat Directions in D7-Brane Moduli Space and 
  Inflationary Model Building'', work in progress.
\bibitem{Kerstan:2011dy}
  M.~Kerstan and T.~Weigand,
  ``The Effective action of D6-branes in N=1 type IIA orientifolds,''
  JHEP {\bf 1106} (2011) 105
  [arXiv:1104.2329 [hep-th]]. \\
  T.~W.~Grimm and D.~V.~Lopes,
  ``The N=1 effective actions of D-branes in Type IIA and IIB orientifolds,''
  Nucl.\ Phys.\ B {\bf 855} (2012) 639
  [arXiv:1104.2328 [hep-th]].
\bibitem{Bohm:1999uk}
  R.~Bohm, H.~Gunther, C.~Herrmann and J.~Louis,
  ``Compactification of type IIB string theory on Calabi-Yau threefolds,''
  Nucl.\ Phys.\ B {\bf 569} (2000) 229 [hep-th/9908007];\\
  T.~W.~Grimm and J.~Louis,
  ``The Effective action of type IIA Calabi-Yau orientifolds,''
  Nucl.\ Phys.\ B {\bf 718} (2005) 153 [hep-th/0412277];\\
  R.~D'Auria, S.~Ferrara and M.~Trigiante,
  ``C - map, very special quaternionic geometry and dual Kahler spaces,''
  Phys.\ Lett.\ B {\bf 587} (2004) 138 [hep-th/0401161];\\
  S.~Alexandrov, B.~Pioline, F.~Saueressig and S.~Vandoren,
  ``D-instantons and twistors,''
  JHEP {\bf 0903} (2009) 044 [arXiv:0812.4219 [hep-th]].
\bibitem{Donagi:2011dv} 
  R.~Donagi and M.~Wijnholt,
  ``Gluing Branes II: Flavour Physics and String Duality,''
  arXiv:1112.4854 [hep-th].
\bibitem{Maharana:2012tu} 
  A.~Maharana and E.~Palti,
  ``Models of Particle Physics from Type IIB String Theory and F-theory: A Review,''
  arXiv:1212.0555 [hep-th].\\
  T.~Weigand,
  ``Lectures on F-theory compactifications and model building,''
  Class.\ Quant.\ Grav.\  {\bf 27}, 214004 (2010)
  [arXiv:1009.3497 [hep-th]].
\bibitem{Aparicio:2008wh}
  L.~Aparicio, D.~G.~Cerdeno and L.~E.~Ib\'a\~nez,
  ``Modulus-dominated SUSY-breaking soft terms in F-theory and their test at 
  LHC,'' JHEP {\bf 0807} (2008) 099 [arXiv:0805.2943 [hep-ph]].
\bibitem{Conlon:2006tj}
  J.~P.~Conlon, D.~Cremades and F.~Quevedo,
  ``Kahler potentials of chiral matter fields for Calabi-Yau string 
  compactifications,'' JHEP {\bf 0701} (2007) 022 [hep-th/0609180].
\bibitem{Beasley:2008dc}
  C.~Beasley, J.~J.~Heckman and C.~Vafa,
  ``GUTs and Exceptional Branes in F-theory - I,''
  JHEP {\bf 0901} (2009) 058
  [arXiv:0802.3391 [hep-th]].
\bibitem{Font:2009gq}
  A.~Font and L.~E.~Ib\'a\~nez,
  ``Matter wave functions and Yukawa couplings in F-theory Grand Unification,''
  JHEP {\bf 0909} (2009) 036 [arXiv:0907.4895 [hep-th]];\\
  J.~P.~Conlon and E.~Palti,
  ``Aspects of Flavour and Supersymmetry in F-theory GUTs,''
  JHEP {\bf 1001} (2010) 029 [arXiv:0910.2413 [hep-th]]; \\
  F.~Marchesano, P.~McGuirk and G.~Shiu,
  ``Chiral matter wavefunctions in warped compactifications,''
  JHEP {\bf 1105} (2011) 090
  [arXiv:1012.2759 [hep-th]].
  
  
\bibitem{Ibanez:1998rf}
  L.~E.~Ib\'a\~nez, C.~Munoz and S.~Rigolin,
  ``Aspect of type I string phenomenology,''
  Nucl.\ Phys.\ B {\bf 553} (1999) 43 [hep-ph/9812397].
\bibitem{Lust:2004cx}
  D.~L\"ust, P.~Mayr, R.~Richter and S.~Stieberger,
  ``Scattering of gauge, matter, and moduli fields from intersecting branes,''
  Nucl.\ Phys.\ B {\bf 696} (2004) 205 [hep-th/0404134];\\
  D.~L\"ust, S.~Reffert and S.~Stieberger,
  ``Flux-induced soft supersymmetry breaking in chiral type IIB orientifolds 
  with D3 / D7-branes,'' Nucl.\ Phys.\ B {\bf 706} (2005) 3 [hep-th/0406092].
\bibitem{Kawano:2011aa} 
  T.~Kawano, Y.~Tsuchiya and T.~Watari,
  ``A Note on Kahler Potential of Charged Matter in F-theory,''
  Phys.\ Lett.\ B {\bf 709}, 254 (2012)
  [arXiv:1112.2987 [hep-th]].
\bibitem{Chacko:2000fn}
  Z.~Chacko and M.~A.~Luty,
  ``Radion mediated supersymmetry breaking,''
  JHEP {\bf 0105} (2001) 067 [hep-ph/0008103];\\
  D.~E.~Kaplan and N.~Weiner,
  ``Radion mediated supersymmetry breaking as a Scherk-Schwarz theory,''
  hep-ph/0108001.
\bibitem{Marti:2001iw}
  D.~Marti and A.~Pomarol,
  ``Supersymmetric theories with compact extra dimensions in N=1 superfields,''
  Phys.\ Rev.\ D {\bf 64} (2001) 105025
  [hep-th/0106256].
\bibitem{Fayet:1975yi}
  P.~Fayet, ``Fermi-Bose Hypersymmetry,''
  Nucl.\ Phys.\ B {\bf 113} (1976) 135.
\bibitem{Strathdee:1986jr}
  J.~A.~Strathdee, ``Extended Poincare Supersymmetry,''
  Int.\ J.\ Mod.\ Phys.\ A {\bf 2} (1987) 273.
\bibitem{Hebecker:2004xx}
  A.~Hebecker and A.~Westphal,
  ``Gauge unification in extra dimensions: Power corrections vs. 
  higher-dimension operators,'' Nucl.\ Phys.\ B {\bf 701} (2004) 273
  [hep-th/0407014].
\bibitem{Blumenhagen:2006xt}
  R.~Blumenhagen, M.~Cveti{\v c} and T.~Weigand,
  ``Spacetime instanton corrections in 4D string vacua: The Seesaw mechanism 
  for D-Brane models,''
  Nucl.\ Phys.\ B {\bf 771} (2007) 113 
  [hep-th/0609191]. \\
  L.~E.~Ib\'a\~nez and A.~M.~Uranga,
  ``Neutrino Majorana Masses from String Theory Instanton Effects,''
  JHEP {\bf 0703} (2007) 052
  [hep-th/0609213]. \\
  M.~Buican, D.~Malyshev, D.~R.~Morrison, H.~Verlinde and M.~Wijnholt,
  ``D-branes at Singularities, Compactification, and Hypercharge,''
  JHEP {\bf 0701} (2007) 107
  [hep-th/0610007]. \\
  L.~E.~Ib\'a\~nez and A.~M.~Uranga,
  ``Instanton induced open string superpotentials and branes at singularities,''
  JHEP {\bf 0802} (2008) 103
  [arXiv:0711.1316 [hep-th]]. \\
  M.~Berg, J.~P.~Conlon, D.~Marsh and L.~T.~Witkowski,
  ``Superpotential de-sequestering in string models,''
  JHEP {\bf 1302} (2013) 018
  [arXiv:1207.1103 [hep-th]].
\bibitem{Breitenlohner:1981sm}
  P.~Breitenlohner and M.~F.~Sohnius,
  ``Matter coupling and nonlinear $\sigma$ models in N=2 supergravity,''
  Nucl.\ Phys.\ B {\bf 187} (1981) 409.
\bibitem{Sohnius:1985qm}
  M.~F.~Sohnius, ``Introducing Supersymmetry,''
  Phys.\ Rept.\  {\bf 128} (1985) 39.
\bibitem{Giddings:2001yu}
  S.~B.~Giddings, S.~Kachru and J.~Polchinski,
  ``Hierarchies from fluxes in string compactifications,''
  Phys.\ Rev.\ D {\bf 66} (2002) 106006 [hep-th/0105097].
\bibitem{Burgess:2006mn}
  C.~P.~Burgess, P.~G.~Camara, S.~P.~de Alwis, S.~B.~Giddings, A.~Maharana, 
  F.~Quevedo and K.~Suruliz,
  ``Warped Supersymmetry Breaking,'' JHEP {\bf 0804} (2008) 053
  [hep-th/0610255].
\bibitem{Cremades:2002cs}
  D.~Cremades, L.~E.~Ib\'a\~nez and F.~Marchesano,
  ``Intersecting brane models of particle physics and the Higgs mechanism,''
  JHEP {\bf 0207} (2002) 022 [hep-th/0203160].
\bibitem{Blumenhagen:2006ci}
  R.~Blumenhagen, M.~Cveti{\v c}, P.~Langacker and G.~Shiu,
  ``Toward realistic intersecting D-brane models,''
  Ann.\ Rev.\ Nucl.\ Part.\ Sci.\  {\bf 55}, 71 (2005)
  [hep-th/0502005].\\
  R.~Blumenhagen, B.~Kors, D.~L\"ust and S.~Stieberger,
  ``Four-dimensional String Compactifications with D-Branes, Orientifolds and Fluxes,''
  Phys.\ Rept.\  {\bf 445} (2007) 1
  [hep-th/0610327]. \\
  L.~E.~Ib\'a\~nez and A.~M.~Uranga,
  ``String theory and particle physics: An introduction to string phenomenology,''
  Cambridge, UK: Univ. Pr. (2012) 673 p
\bibitem{Cvetic:2001nr}
  M.~Cveti{\v c}, G.~Shiu and A.~M.~Uranga,
  ``Chiral four-dimensional N=1 supersymmetric type 2A orientifolds from intersecting D6 branes,''
  Nucl.\ Phys.\ B {\bf 615} (2001) 3
  [hep-th/0107166].
\bibitem{Honecker:2004kb} 
  G.~Honecker and T.~Ott,
  ``Getting just the supersymmetric standard model at intersecting branes on the Z(6) orientifold,''
  Phys.\ Rev.\ D {\bf 70}, 126010 (2004)
  [Erratum-ibid.\ D {\bf 71}, 069902 (2005)]
  [hep-th/0404055].
\bibitem{Blumenhagen:2000wh}
  R.~Blumenhagen, L.~Goerlich, B.~Kors and D.~L\"ust,
  ``Noncommutative compactifications of type I strings on tori with magnetic background flux,''
  JHEP {\bf 0010} (2000) 006
  [hep-th/0007024]. \\
  G.~Aldazabal, S.~Franco, L.~E.~Ib\'a\~nez, R.~Rabadan and A.~M.~Uranga,
  ``Intersecting brane worlds,''
  JHEP {\bf 0102} (2001) 047
  [hep-ph/0011132].
\bibitem{Ibanez:2001nd}
  L.~E.~Ib\'a\~nez, F.~Marchesano and R.~Rabadan,
  ``Getting just the standard model at intersecting branes,''
  JHEP {\bf 0111} (2001) 002 [hep-th/0105155].
\bibitem{Blumenhagen:2001te}
  R.~Blumenhagen, B.~Kors, D.~L\"ust and T.~Ott,
  ``The standard model from stable intersecting brane world orbifolds,''
  Nucl.\ Phys.\ B {\bf 616} (2001) 3
  [hep-th/0107138].
\bibitem{Kallosh:2001tm}
  R.~Kallosh, ``N=2 supersymmetry and de Sitter space,''
  hep-th/0109168;\\
  R.~Kallosh and A.~D.~Linde, ``P term, D term and F term inflation,''
  JCAP {\bf 0310} (2003) 008 [hep-th/0306058].
\bibitem{ftermdecoupling}
  E.~A.~Mirabelli and M.~E.~Peskin,
  ``Transmission of supersymmetry breaking from a four-dimensional boundary,''
  Phys.\ Rev.\ D {\bf 58}, 065002 (1998)
  [hep-th/9712214].\\
  C.~Bouchart, A.~Knochel and G.~Moreau,
  ``Discriminating 4D supersymmetry from its 5D warped version,''
  Phys.\ Rev.\ D {\bf 84}, 015016 (2011)
  [arXiv:1101.0634 [hep-ph]].
\bibitem{Anastasopoulos:2011kr}
  P.~Anastasopoulos, I.~Antoniadis, K.~Benakli, M.~D.~Goodsell and A.~Vichi,
  ``One-loop adjoint masses for non-supersymmetric intersecting branes,''
  JHEP {\bf 1108} (2011) 120
  [arXiv:1105.0591 [hep-th]].
\bibitem{Berkooz:1996km}
  M.~Berkooz, M.~R.~Douglas and R.~G.~Leigh,
  ``Branes intersecting at angles,''
  Nucl.\ Phys.\ B {\bf 480} (1996) 265
  [hep-th/9606139].
\bibitem{Okada:1990vk} 
  Y.~Okada, M.~Yamaguchi and T.~Yanagida,
  ``Upper bound of the lightest Higgs boson mass in the minimal supersymmetric standard model,''
  Prog.\ Theor.\ Phys.\  {\bf 85}, 1 (1991).\\
  Y.~Okada, M.~Yamaguchi and T.~Yanagida,
  ``Renormalization group analysis on the Higgs mass in the softly broken 
  supersymmetric standard model,''
  Phys.\ Lett.\ B {\bf 262}, 54 (1991).
\bibitem{Cornwall:2012ea} 
  A.~Kusenko, V.~Kuzmin and I.~I.~Tkachev,
  Phys.\ Lett.\ B {\bf 432}, 361 (1998)
  [hep-ph/9801405].\\
  J.~M.~Cornwall, A.~Kusenko, L.~Pearce and R.~D.~Peccei,
  arXiv:1210.6433 [hep-ph].
\bibitem{pdgtop}
 J. Beringer et al. (Particle Data Group), 
 PR D86, 010001 (2012)\\ 
 (URL: http://pdg.lbl.gov)
\bibitem{Dobado:2002jz} 
  A.~Dobado, M.~J.~Herrero, W.~Hollik and S.~Penaranda,
  ``Selfinteractions of the lightest MSSM Higgs boson in the large 
  pseudoscalar mass limit,''
  Phys.\ Rev.\ D {\bf 66}, 095016 (2002) [hep-ph/0208014].
\bibitem{higherdsuperspace}
  N.~Arkani-Hamed, T.~Gregoire and J.~G.~Wacker,
  ``Higher dimensional supersymmetry in 4-D superspace,''
  JHEP {\bf 0203}, 055 (2002)
  [hep-th/0101233].
  A.~Hebecker,
  ``5-D superYang-Mills theory in 4-D superspace, superfield brane operators, and applications to orbifold GUTs,''
  Nucl.\ Phys.\ B {\bf 632}, 101 (2002)
  [hep-ph/0112230].
\bibitem{p-term}
  E.~Halyo,
  hep-th/0405269.
\\
  R.~Kallosh and A.~D.~Linde,
  ``P term, D term and F term inflation,''
  JCAP {\bf 0310}, 008 (2003)
  [hep-th/0306058].
\bibitem{Guth:1982ec} 
  A.~H.~Guth and S.~Y.~Pi,
  ``Fluctuations in the New Inflationary Universe,''
  Phys.\ Rev.\ Lett.\  {\bf 49}, 1110 (1982);\\
  A.~A.~Starobinsky,
  ``Dynamics of Phase Transition in the New Inflationary Universe Scenario 
  and Generation of Perturbations,'' Phys.\ Lett.\ B {\bf 117} (1982) 175;\\
  A.~Vilenkin,
  ``Topological inflation,''
  Phys.\ Rev.\ Lett.\  {\bf 72} (1994) 3137
  [hep-th/9402085];\\
  L.~Boubekeur and D.~.H.~Lyth,
  ``Hilltop inflation,''
  JCAP {\bf 0507} (2005) 010
  [hep-ph/0502047].
\bibitem{Vilenkin:1982wt}
  A.~Vilenkin, L.~H.~Ford
  ``Gravitational Effects upon Cosmological Phase Transitions,''
  Phys.\ Rev.\ D {\bf 26} (1982) 1231.
\bibitem{Starobinsky:1986fx}
  A.~A.~Starobinsky,
  ``Stochastic De Sitter (inflationary) Stage In The Early Universe,''
  In *De Vega, H.j. ( Ed.), Sanchez, N. ( Ed.): Field Theory, Quantum 
  Gravity and Strings*, 107-126;\\
  M.~Sasaki, H.~Suzuki, K.~Yamamoto, J.~'i.~Yokoyama,
  ``Superexpansionary divergence: Breakdown of perturbative quantum field 
  theory in space-time with accelerated expansion,''
  Class.\ Quant.\ Grav.\  {\bf 10} (1993) L55;\\
  A.~A.~Starobinsky, J.~Yokoyama,
  ``Equilibrium state of a selfinteracting scalar field in the De Sitter 
  background,'' Phys.\ Rev.\ D {\bf 50} (1994) 6357
  [astro-ph/9407016].
\bibitem{Lebedev:2012sy}
  O.~Lebedev and A.~Westphal,
  ``Metastable Electroweak Vacuum: Implications for Inflation,''
  Phys.\ Lett.\ B {\bf 719} (2013) 415
  [arXiv:1210.6987 [hep-ph]].

\bibitem{Riotto:1995am}
  A.~Riotto and E.~Roulet,
  ``Vacuum decay along supersymmetric flat directions,''
  Phys.\ Lett.\ B {\bf 377} (1996) 60
  [hep-ph/9512401];\\
  A.~Kusenko, P.~Langacker and G.~Segre,
  ``Phase transitions and vacuum tunneling into charge and color breaking minima in the MSSM,''
  Phys.\ Rev.\ D {\bf 54} (1996) 5824
  [hep-ph/9602414];\\
  S.~A.~Abel, C.~-S.~Chu, J.~Jaeckel and V.~V.~Khoze,
  ``SUSY breaking by a metastable ground state: Why the early universe 
  preferred the non-supersymmetric vacuum,''
  JHEP {\bf 0701} (2007) 089
  [hep-th/0610334];\\
  J.~R.~Ellis, J.~Giedt, O.~Lebedev, K.~Olive and M.~Srednicki,
  ``Against Tachyophobia,''
  Phys.\ Rev.\ D {\bf 78} (2008) 075006
  [arXiv:0806.3648 [hep-ph]].

\bibitem{Nath:2012nh}
  P.~Nath,
  ``Higgs Physics and Supersymmetry,''
  Int.\ J.\ Mod.\ Phys.\ A {\bf 27} (2012) 1230029
  [arXiv:1210.0520 [hep-ph]] and arXiv:1302.1863 [hep-ph].\\
  P.~Nath,
  ``Perspectives on Higgs Boson and Supersymmetry,''
  arXiv:1302.1863 [hep-ph].
\bibitem{Arbey:2013jla}
  A.~Arbey, M.~Battaglia and F.~Mahmoudi,
  ``Supersymmetric Heavy Higgs Bosons at the LHC,''
  arXiv:1303.7450 [hep-ph];\\
  A.~Djouadi and J.~Quevillon,
  ``The MSSM Higgs sector at a high $M_{SUSY}$: reopening the low tan$\beta$ regime and the search for heavy Higgsses,''
  arXiv:1304.1787 [hep-ph].
\bibitem{Arvanitaki:2012ps}
  A.~Arvanitaki, N.~Craig, S.~Dimopoulos and G.~Villadoro,
  ``Mini-Split,''\\
  arXiv:1210.0555 [hep-ph];
\bibitem{Chatzistavrakidis:2012bb} 
  A.~Chatzistavrakidis, E.~Erfani, H.~P.~Nilles and I.~Zavala,
  ``Axiology,''
  JCAP {\bf 1209}, 006 (2012)
  [arXiv:1207.1128 [hep-ph]].
\bibitem{Davies:2011mp}
  R.~Davies, J.~March-Russell and M.~McCullough,
  ``A Supersymmetric One Higgs Doublet Model,''
  JHEP {\bf 1104} (2011) 108 [arXiv:1103.1647 [hep-ph]].
\bibitem{Higaki:2012ar} 
  T.~Higaki and F.~Takahashi,
  ``Dark Radiation and Dark Matter in Large Volume Compactifications,''
  JHEP {\bf 1211}, 125 (2012)
  [arXiv:1208.3563 [hep-ph]].\\
  M.~Cicoli, J.~P.~Conlon and F.~Quevedo,
  ``Dark Radiation in LARGE Volume Models,''
  Phys.\ Rev.\ D {\bf 87}, 043520 (2013)
  [arXiv:1208.3562 [hep-ph]].\\
  T.~Higaki, K.~Kamada and F.~Takahashi,
  ``Higgs, Moduli Problem, Baryogenesis and Large Volume Compactifications,''
  JHEP {\bf 1209}, 043 (2012)
  [arXiv:1207.2771 [hep-ph]].



\end{thebibliography}
\end{document}